\documentclass[11pt,letter]{article}
\pdfoutput=1
\usepackage{jheppub}
\allowdisplaybreaks
\usepackage{xcolor}
\usepackage{amsmath}
\usepackage{amssymb}
\usepackage{graphicx}
\usepackage{esint}
\usepackage{slashed}
\usepackage[english]{babel}
\usepackage{hyperref}
\usepackage[normalem]{ulem}
\usepackage{comment}

\definecolor{kjkblue}{rgb}{0.39, 0.589, 0.6914}

\begin{document}

\hfill \preprint{FERMILAB-PUB-21-200-T, MI-TH-218}

\title{Light, Long-Lived $B-L$ Gauge and Higgs Bosons at the DUNE Near Detector}

\author[a]{P. S. Bhupal Dev,}
\author[b]{Bhaskar Dutta,}
\author[c]{Kevin J. Kelly,}
\author[d]{Rabindra N. Mohapatra,}
\author[e,a]{Yongchao Zhang}

\affiliation[a]{Department of Physics and McDonnell Center for the Space Sciences, Washington University, St. Louis, MO 63130, USA}
\affiliation[b]{Mitchell Institute for Fundamental Physics and Astronomy,
Department of Physics and Astronomy,
Texas A\&M University, College Station, TX 77845, USA}
\affiliation[c]{Theoretical Physics Department, Fermi National Accelerator Laboratory, P. O. Box 500, Batavia, IL 60510, USA }
\affiliation[d]{Maryland Center for Fundamental Physics, Department of Physics, University of Maryland, College Park, MD 20742, USA}
\affiliation[e]{School of Physics, Southeast University, Nanjing 211189, China}

\emailAdd{bdev@wustl.edu}
\emailAdd{dutta@physics.tamu.edu}
\emailAdd{kkelly12@fnal.gov}
\emailAdd{rmohapat@umd.edu}
\emailAdd{zhangyongchao@seu.edu.cn}

\date{\today}

\abstract{
The low-energy $U(1)_{B-L}$ gauge symmetry is well-motivated as part of beyond Standard Model physics related to neutrino mass generation. We show that a light $B-L$ gauge boson $Z{'}$ and the associated $U(1)_{B-L}$-breaking scalar $\varphi$ can both be effectively searched for at high-intensity facilities such as the near detector complex of the Deep Underground Neutrino Experiment (DUNE). Without the scalar $\varphi$, the $Z{'}$ can be probed at DUNE up to mass of 1 GeV, with the corresponding gauge coupling $g_{BL}$ as low as $10^{-9}$. In the presence of the scalar $\varphi$ with gauge coupling to $Z{'}$, the DUNE capability of discovering the gauge boson $Z{'}$ can be significantly improved, even by one order of magnitude in $g_{BL}$, due to additional production from the decay $\varphi \to Z{'}Z{'}$. The DUNE sensitivity is largely complementary to other long-lived $Z{'}$ searches at beam-dump facilities such as FASER and SHiP, as well as astrophysical and cosmological probes. On the other hand, the prospects of detecting $\varphi$ itself at DUNE are to some extent weakened in presence of $Z{'}$, compared to the case without the gauge interaction.
}

\maketitle

\section{Introduction}\label{sec:Introduction}

The $U(1)_{B-L}$ gauge symmetry is a well-motivated extension~\cite{Davidson:1978pm, Marshak:1979fm}  of the Standard Model (SM), which provides a natural framework to account for the tiny neutrino masses via the type-I seesaw mechanism~\cite{Minkowski:1977sc,Mohapatra:1979ia,Yanagida:1979as,GellMann:1980vs, Glashow:1979nm}. 
It may also accommodate a light dark matter (DM) particle which interacts with the SM particles through the scalar or gauge portal of $U(1)_{B-L}$~\cite{Mohapatra:2019ysk}. In this paper, we focus on the class of $B-L$ extensions of SM, where $B-L$ does not contribute to electric charge so that its gauge coupling $g_{BL}$ can be arbitrarily small making the associated gauge boson $Z'$ very light with mass in the sub-GeV range~\cite{Brehmer:2015cia, Dev:2016dja, Chauhan:2018uuy}. This is especially true if the $U(1)_{B-L}$-breaking scale is low, e.g., below the GeV-scale.
In this case, we can expect the corresponding $U(1)_{B-L}$-breaking scalar (denoted here by $\varphi$) to be also light. Both $Z'$ and $\varphi$ are essential ingredients of the $U(1)_{B-L}$ extension, and play important role in connecting DM to the SM, in DM phenomenology~\cite{Okada:2012sg,Kaneta:2016vkq,Klasen:2016qux,  Mohapatra:2019ysk, Heeba:2019jho, Mohapatra:2020bze, Borah:2020wyc}, and in explaining the observed light neutrino masses~\cite{Wetterich:1981bx, Buchmuller:1991ce, Emam:2007dy, Basso:2008iv, Perez:2009mu, Basso:2010jm, Heeck:2014zfa}. Since the seesaw mechanism does not depend directly on the gauge coupling value, such low gauge coupling models can still accommodate the seesaw mechanism. For phenomenological exploration of this class of $B-L$ extensions at a high mass scale and larger gauge coupling range, see, e.g., Refs.~\cite{Khalil:2006yi,Huitu:2008gf,Accomando:2016sge, Accomando:2017qcs, Dev:2017xry, Alioli:2017nzr}.


If kinematically allowed, both $\varphi$ and $Z'$ can be produced and detected in the high-intensity experiments such as the Deep Underground Neutrino Experiment (DUNE)~\cite{Berryman:2019dme}. One or both of these particles may be relatively long-lived and be able to travel from the DUNE target to its near detector complex where it decays in a striking fashion. It may then be possible to search for signatures of this class of beyond Standard Model (BSM) physics in such experiments.
Our goal in this work is to determine how these gauge bosons, scalar bosons, and their interplay can be searched for in the DUNE experiment at Fermilab.
In particular, we will focus on the following three scenarios: (i) pure $Z'$ case with the scalar $\varphi$ decoupled, (ii) the effect of $\varphi \to Z'Z'$ on the prospects of $\varphi$ at DUNE, and (iii) improvement of the DUNE sensitivity of $Z'$ boson due to the scalar source $\varphi \to Z'Z'$.

The $Z'$ boson in the $U(1)_{B-L}$ model couples to all the SM quarks and leptons and thus can be produced from meson decays such as $\pi \to \gamma+ Z'$ and $\eta\to \gamma+Z'$ as well as other hadronic decays where the $\pi^0$ and $\eta$ etc are final state particles in the $pp$ collision~\cite{Batell:2009di}. The proton-proton bremsstrahlung process is also very important, in particular when the $Z'$ boson is heavier than the $\eta$ meson~\cite{Blumlein:2013cua, deNiverville:2016rqh} (cf. Fig.~\ref{fig:flux:Zprime}). After being produced, the $Z'$ boson may decay via $Z' \to e^+ e^-,\, \mu^+ \mu^- ,\, \nu\bar\nu$ and hadrons such as $\pi\pi\pi$ and $KK$~\cite{Buschmann:2015awa,Ilten:2018crw,Bauer:2018onh}, depending on its mass,  and in the small gauge coupling limit can lead to displaced vertices with decays in the DUNE near detector hundreds of meters away. The observation of such displaced vertices may provide the key signatures of the light $Z'$ boson. An analogous analysis of a leptonic gauge boson (e.g. one of $U(1)_{L_\mu - L_\tau}$) has been performed in Ref.~\cite{Berryman:2019dme}. Such gauge bosons can lead to anomalous neutrino-electron scattering and neutrino trident events, which have been explored in Refs.~\cite{Ballett:2019xoj,Altmannshofer:2019zhy}.
We will perform a thorough study of the $B-L$ $Z'$ gauge boson and determine the DUNE near detector sensitivity. It turns out to be qualitatively different from the leptonic $Z'$ boson, as presented in Fig.~\ref{fig:prospect:PureZprime}.

As for the light scalar $\varphi$, if it mixes with the SM Higgs boson, it will obtain  loop-level flavor-changing neutral current (FCNC) couplings to the SM quarks. As a result, it can be produced from the loop-level FCNC meson decays such as $K \to \pi + \varphi$ at DUNE~\cite{Batell:2009jf}, and then  decay into the light SM particles $\varphi \to e^+ e^-,\, \mu^+ \mu^-,\, \pi \pi,\, \gamma\gamma$, with the decay also induced by the mixing of $\varphi$ with the SM Higgs. The sensitivity of a light scalar $\varphi$ without the $Z'$ boson at DUNE has been studied in Ref.~\cite{Berryman:2019dme}.

In the $U(1)_{B-L}$ model, there is the gauge coupling $\varphi Z'Z'$, This will induce the extra decay channel $\varphi \to Z' Z'$,
which could have multiple effects on the prospects of $\varphi$ and $Z'$ at DUNE:
\begin{itemize}
    \item This new decay channel of $\varphi$ will produce more $Z'$ bosons at DUNE, compared to the pure $Z'$ case, even by a factor of $10^5$ (see Fig.~\ref{fig:CombinedExplanation}). As a result, the $Z'$ can be probed at DUNE with a smaller gauge coupling $g_{BL}$, with an improvement factor up to 45 (see Fig.~\ref{fig:prospect:CombinedZPrimeScalar}). This is only possible for a $Z'$ mass $m_{Z'} \lesssim 200$ MeV. This is because we require $m_{\varphi} > 2m_{Z'}$, and $\varphi$ heavier than ${\sim}400$ MeV cannot be produced from the meson decay $K \to \pi + \varphi$ at DUNE.

    \item The gauge portal scalar decay will not only change the branching fractions of $\varphi$ (cf. Fig.~\ref{fig:BRs}), but also shorten its lifetime. This will have some effects on the search for $\varphi$ at DUNE, depending on the BR of $\varphi$ decaying into SM particles, i.e. ${\rm BR} (\varphi \to {\rm SM})$. As presented in Fig.~\ref{fig:scalar:varyBr}, the effect of $\varphi \to Z'Z'$ is most significant when $\varphi$ is relatively heavy and the mixing angle $\sin\vartheta$ is relatively large.

    \item The simultaneous existence of $\varphi$ and $Z'$ will also induce the associated production of $\varphi$ and $Z'$ at DUNE, for instance from the meson decay $K \to \pi + Z' + \varphi$. However, no matter whether the scalar $\varphi$ is emitted from the $Z'$ boson line or from couplings to mesons, such decays will always be highly suppressed by either $g_{BL}^4$ or $\sin^2\vartheta g_{BL}^2$, and can thus be neglected at DUNE.
\end{itemize}





The rest of this paper is organized as follows. First, we provide a brief review of the model in Section~\ref{sec:Model}. Then, we begin the discussion of search strategies in Section~\ref{sec:Search} with a brief discussion of the experimental setup and how proton-beam experiments (including the DUNE beamline) are well-suited for searches of these types of models. We divide the experimental search for this model by taking the following approach: first, we discuss in Section~\ref{subsec:PureZPrime} how the $Z'$ gauge boson of $U(1)_{B-L}$ can be searched for in DUNE making agnostic assumptions about the existence of the associated scalar $\varphi$. Then, in Section~\ref{subsec:PureScalar},
we will demonstrate that searches for $\varphi$ in accelerator beam environments, including DUNE, are weakened in the presence of $Z'$, where the produced $\varphi$ can decay into $Z'$ instead of SM particles. We will also revisit the $Z'$ search strategy and discuss how additional fluxes sourced by $\varphi$ (including a brief discussion of associated production of both $\varphi$ and $Z'$) decays can be detected in Section~\ref{subsec:CombinedSearch}. This improves the capability of discovering the gauge boson $Z'$. Finally, we conclude in Section~\ref{sec:conclusion}.
Some details of the $\varphi$ decay calculations are relegated to Appendix~\ref{app:A}. Subdominant contributions to $Z'$ flux at DUNE are collected in Appendix~\ref{appendix:ZPrimeFluxes}, and more $Z'$ limits can be found in Appendix~\ref{appendix:Zprimelimit}.

\section{The Model}
\label{sec:Model}

We consider the minimal $U(1)_{B-L}$ model based on the gauge group $SU(2)_L \times U(1)_Y \times U(1)_{B-L}$. For the purpose of anomaly cancellation, three right-handed neutrinos (RHNs) $N_i$ (with $i=1,\, 2,\, 3$) are naturally introduced. To break the $U(1)_{B-L}$ gauge symmetry, a complex singlet scalar field $\phi$ is introduced, which carries two units of $B-L$ charge. When the $\phi$-field  develops a non-vanishing vacuum expectation value (VEV) $\langle \phi\rangle = v_{BL}/2\sqrt{2}$, the $U(1)_{B-L}$ gauge symmetry is spontaneously broken, which generates the $Z'$ boson mass of $m_{Z'} = g_{BL} v_{BL}$, where we have defined the covariant derivative as $D_\mu \phi=\left[\partial_\mu - 2ig_{BL} Z'_{\mu} \right] \phi$. After the symmetry breaking, expanding the field around its VEV, we obtain
\begin{eqnarray}
 \phi = \frac{1}{\sqrt{2}}\left(\frac{1}{2} v_{BL} +\varphi+i\chi\right) \,.
\end{eqnarray}
  The CP-odd component $\chi$ is ``eaten" by the $Z'$ boson, while the CP-even component $\varphi$ is left as a physical scalar field. In presence of the $(H^\dagger H) (\phi^\dagger \phi)$ term in the scalar potential ($H$ is the SM Higgs doublet), the scalar $\varphi$ mixes with the SM Higgs $h$, with a mixing angle $\sin\vartheta$.

Given the Yukawa coupling
\begin{eqnarray}
{\cal L}_Y = - y_N \phi \overline{N^C} N \,,
\end{eqnarray}
with $C$ standing here for charge conjugate, the RHNs obtain masses $m_N = y_N v_{BL}/\sqrt2$, which can be used to generate the tiny active neutrino masses via type-I seesaw mechanism~\cite{Minkowski:1977sc,Mohapatra:1979ia,Yanagida:1979as,GellMann:1980vs, Glashow:1979nm}. For simplicity, we will assume the RHN masses $m_N > m_{Z'}/2$ and $m_N > m_\varphi/2$ such that the decays $Z' \to NN$ and $\varphi \to NN$ are kinematically forbidden, and we can neglect the contributions of RHNs to the DUNE prospects of $Z'$ and $\varphi$ in this paper. We also assume for simplicity that there is no $Z - Z'$ mixing throughout this paper, which would otherwise potentially affect the decay branching ratios (BRs) of the $Z'$ boson. A small, loop-induced kinetic mixing between $Z^\prime$ and the SM $Z$ will arise, on the order of $g_{BL} g/(16\pi^2)$. Since we are interested in $g_{BL} \ll 1$, this mixing is always very small, and we will disregard it for the remainder of this work. We are also going to focus on the mass regime MeV $\lesssim m_{Z'} \lesssim$ GeV -- whether such small gauge couplings and light gauge boson masses are natural or fine-tuned requires a detailed analysis of the scalar potential, which is beyond the scope of this manuscript, and we just take a more phenomenological approach.

Then there are  only four free phenomenological parameters in the minimal $U(1)_{B-L}$ model, i.e.
\begin{eqnarray}
\label{eqn:parameters}
m_{Z'} \,, \quad m_{\varphi} \,, \quad
g_{BL} \,, \quad \sin\vartheta \,,
\end{eqnarray}
which we study to investigate the DUNE prospects. To produce the $Z'$ boson and/or the $\varphi$ scalar at the DUNE experiment, their masses $m_{Z'}$ and $m_\varphi$ have to be at or below the GeV-scale. As we will show in Figs.~\ref{fig:prospect:PureZprime} and \ref{fig:scalar:varyBr} respectively, the gauge coupling $g_{BL}$ and the mixing $\sin\vartheta$ have to be sufficiently smaller than one to satisfy the current experimental constraints.

\begin{figure}[!t]
  \centering
  \includegraphics[width=0.65\textwidth]{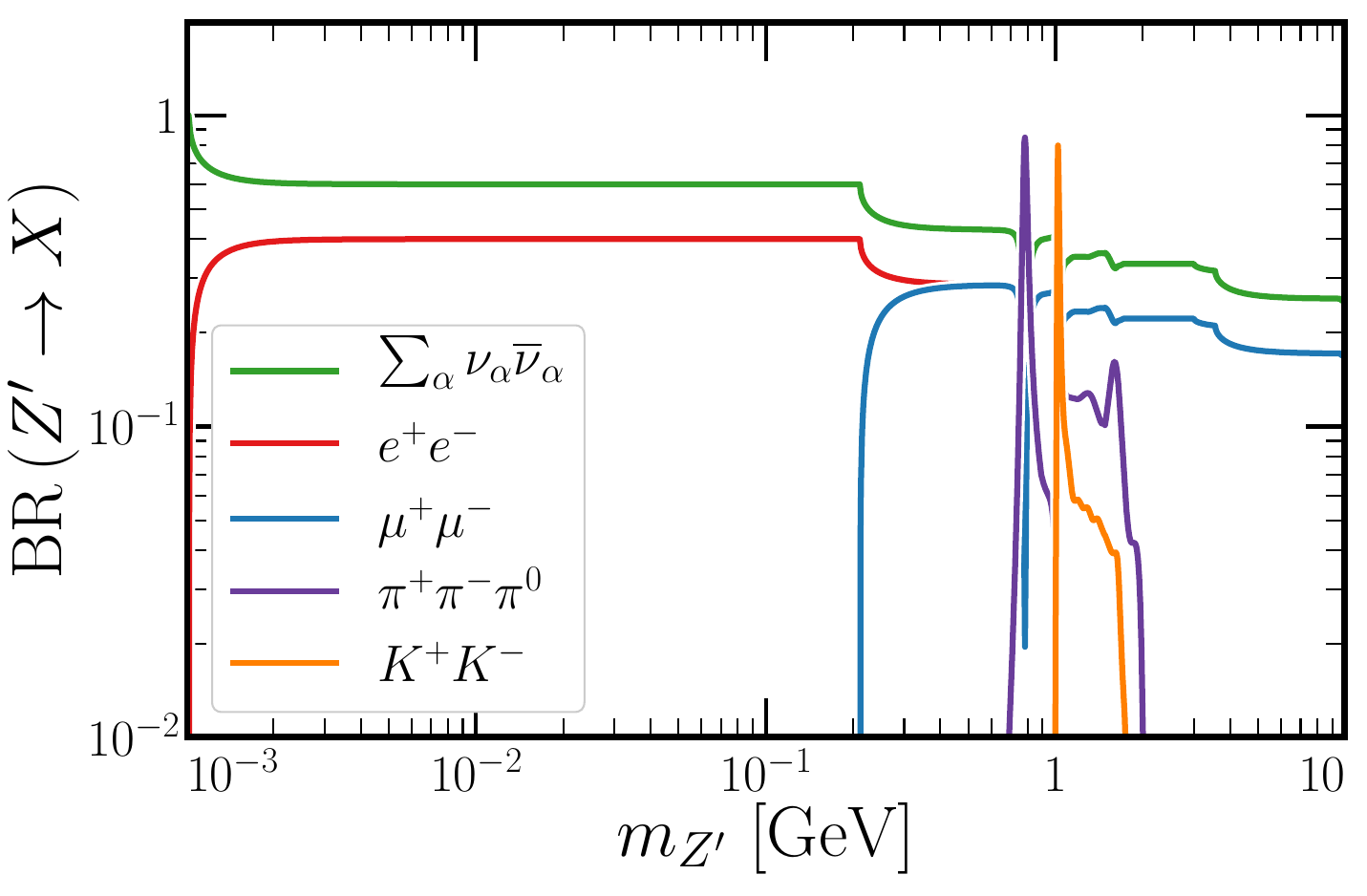}
  \caption{Decay BRs of the $Z'$ in $U(1)_{B-L}$ model into $\nu\bar\nu$, $e^+ e^-$, $\mu^+ \mu^+ $ and the dominant hadronic channels. Data are obtained using the code {\sc darkcast}~\cite{Ilten:2018crw}.
  }
  \label{fig:Zprimedecay}
\end{figure}

Neglecting the RHN channel, the gauge boson $Z'$ decays predominantly into pairs of SM fermions (and mesons), and we can express the decay widths of $Z'$ into SM fermions as~\cite{Ilten:2018crw}
\begin{eqnarray}
\Gamma (Z' \to \bar{f} f) = \frac{m_{Z'} g_{BL}^2 N^C_f S_f}{12\pi}
\left( 1 + \frac{2m_f^2}{m_{Z'}^2} \right)
\sqrt{ 1 - \frac{4m_f^2}{m_{Z'}^2} } \,,
\end{eqnarray}
with $N^C_f$ being the color factor ($3$ for quarks and $1$ for leptons), and the symmetry factor $S_f = 1$ for quarks and charged leptons and $1/2$ for neutrinos. Decay widths into hadrons from the quark hadronization are more complicated, but can be expressed using the $R$-ratio (see, e.g., Refs.~\cite{Buschmann:2015awa,Ilten:2018crw,Bauer:2018onh} for further discussion). In practice, we use the {\sc darkcast} code from Ref.~\cite{Ilten:2018crw} to determine the $Z'$ lifetime and BR into visible final states accessible in the DUNE near detector. These final states include $e^+ e^-$, $\mu^+ \mu^-$, and hadronic ones, dominated by $\pi^+ \pi^- \pi^0$ and $K^+ K^-$ states (but not the two-pion final state, which is forbidden by $G$-parity conservation in the absence of isospin-breaking terms); Fig.~\ref{fig:Zprimedecay} presents these BRs as a function of $m_{Z'}$.

Through mixing with the SM Higgs, the scalar $\varphi$ decays into the SM leptons, pion pairs and two photons via the SM $W$ boson and charged fermion loops. The corresponding partial widths are respectively
\begin{eqnarray}
\label{eqn:phiff}
\Gamma (\varphi \to \ell^+ \ell^-) & = &
\frac{G_F m_\varphi m_\ell^2\sin^2\vartheta}{4\sqrt2 \pi}
\left( 1 - \frac{4m_\ell^2}{m_\varphi^2} \right)^{3/2} \,, \\
\label{eqn:phipipi}
\Gamma (\varphi \to \pi^+ \pi^-) & = & 2 \Gamma (\varphi \to \pi^0 \pi^0) =
\frac{G_F m_\varphi^3 \sin^2\vartheta}{8\sqrt2 \pi}
\left( 1 - \frac{4m_\pi^2}{m_\varphi^2} \right)^{1/2}
\left| G (m_\varphi^2) \right|^2 \,, \\
\label{eqn:phiAA}
\Gamma (\varphi \to \gamma \gamma ) & = &
\frac{G_F \alpha^2 m_\varphi^3 \sin^2\vartheta}{128 \sqrt2 \pi^3}
\left| \sum_f N_f^C Q_f^2 A_{1/2} (\tau_f) + A_{1} (\tau_W) \right|^2 \,,
\end{eqnarray}
where $G_F$ is the Fermi constant, $\alpha\equiv e^2/4\pi$ is the fine-structure constant, and $Q_f$ is the charge of the fermion $f$ in units of the proton charge $e$. In Eq.~\eqref{eqn:phipipi}, $G (m_\varphi^2)$ is a dimensionless transition amplitude defined in Appendix~\ref{app:A}. Similarly in Eq.~\eqref{eqn:phiAA}, $A_{1/2}(\tau_f)$ and $A_1(\tau_W)$ respectively (with $\tau_X=m_\varphi^2/4m_X^2$) are standard fermion and $W$ loop functions for the Higgs decay, also given in Appendix~\ref{app:A} for completeness.

As a result of the gauge coupling of $\varphi$ to $Z'$ bosons, if kinematically allowed, we have also the decay channel
\begin{eqnarray}
\label{eqn:phiZpZp}
\Gamma (\varphi \to Z'Z') & = & \frac{g_{BL}^2 m_\varphi^3}{32 \pi m_{Z'}^2}
\left( 1- \frac{4m_{Z'}^2}{m_\varphi^2} \right)^{1/2}
\left( 1- \frac{4m_{Z'}^2}{m_\varphi^2} + \frac{12m_{Z'}^4}{m_\varphi^4} \right) \,.
\end{eqnarray}

\begin{figure}[!t]
  \centering
  \includegraphics[width=0.6\textwidth]{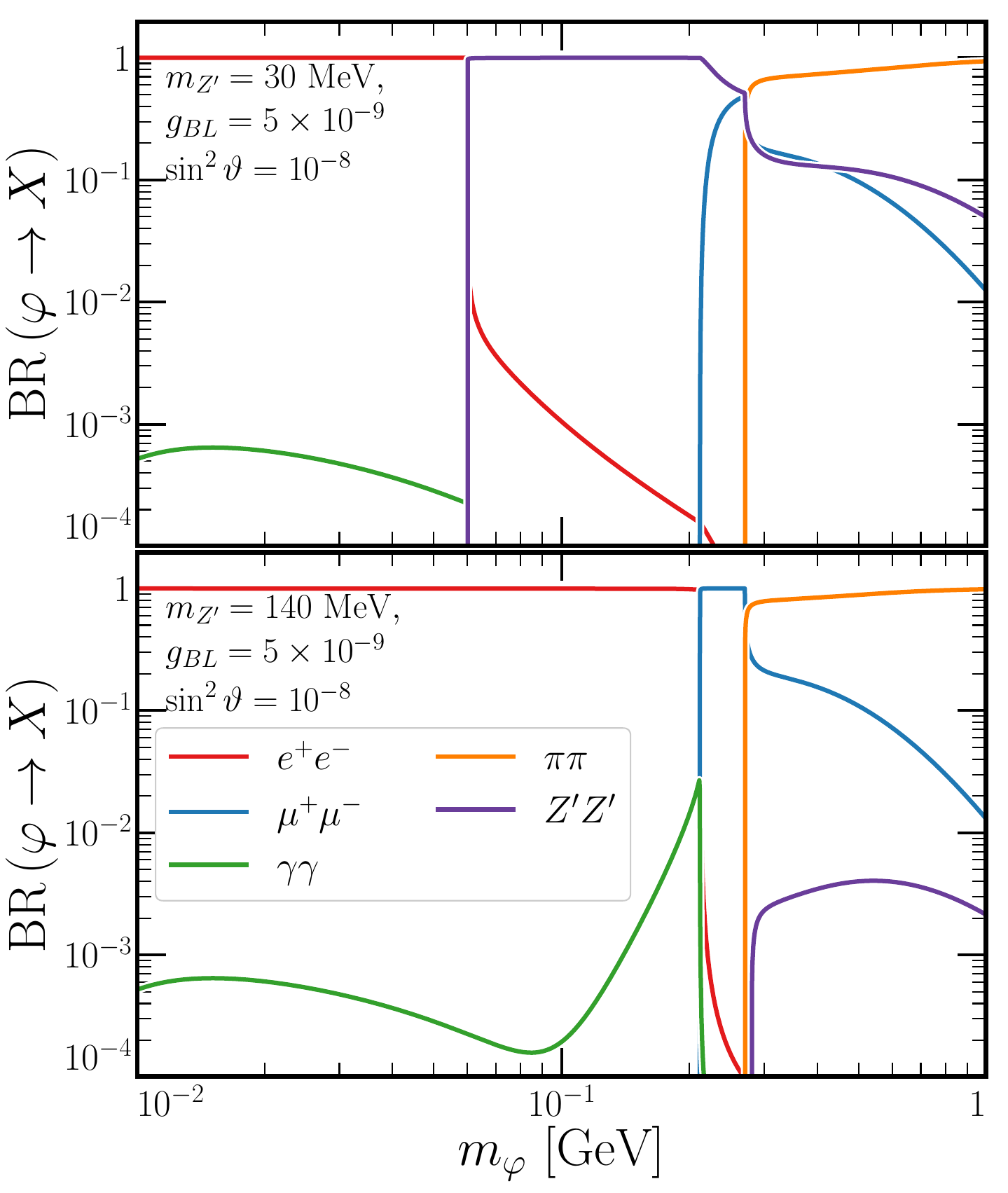}
  \caption{BRs of the scalar $\varphi$ in the $U(1)_{B-L}$ model into $ee$, $\mu\mu$, $\gamma\gamma$, $\pi\pi$ (including both $\pi^0 \pi^0$ and $\pi^+\pi^-$) and $Z'Z'$, as a function of the scalar mass. In both panels, we take $g_{BL} = 5 \times 10^{-9}$ and $\sin^2\vartheta = 10^{-8}$. The top (bottom) panel assumes $m_{Z'} =$ 30 MeV ($140$ MeV).}
  \label{fig:BRs}
\end{figure}

As shown in Eqs.~(\ref{eqn:phiff}) - (\ref{eqn:phiAA}), all the decay channels of $\varphi$ into the SM particles are universally proportional to the mixing angle $\sin\vartheta$, while the $Z'$ boson channel in Eq.~(\ref{eqn:phiZpZp}) is dictated by the gauge coupling $g_{BL}$. In the limit of $g_{BL} \to 0$ or $m_\varphi < 2m_{Z'}$, the scalar $\varphi$ only decays into the SM particles, which is equivalent to the case of singlet extension of the SM. In presence of  the $U(1)_{B-L}$ gauge couplings and the $Z'$ boson, the decay products of $\varphi$ are changed dramatically, depending on the values of $m_{Z'}$, $\sin\vartheta$ and $g_{BL}$. The BRs of $\varphi$ for the following two Benchmark Points (BP) of $m_{Z'}$ and $g_{BL}$ are shown in Fig.~\ref{fig:BRs}:
\begin{eqnarray}
      \label{eqn:BP1}
      {\rm BP1}: && m_{Z'} = 30 \; {\rm MeV} \,, \quad
      g_{BL} = 5 \times 10^{-9} \,, \\
      \label{eqn:BP2}
      {\rm BP2}: && m_{Z'} = 140 \; {\rm MeV} \,, \quad
      g_{BL} = 5 \times 10^{-9} \,,
\end{eqnarray}
which correspond to the two stars in Fig.~\ref{fig:prospect:CombinedZPrimeScalar}.

\section{Searches for Scalars and Gauge Bosons at DUNE}
\label{sec:Search}

The Fermilab DUNE facility~\cite{Acciarri:2015uup, Acciarri:2016crz} uses a 120 GeV proton beam\footnote{Different beam configurations, and in particular, the possibility of using proton beam with 80 GeV energy, have also been considered by the collaboration~\cite{Acciarri:2016crz}. Appendix A of Ref.~\cite{Berryman:2019dme} explored the production rates of various SM mesons for both 80 GeV and 120 GeV scenarios and found that the differences are relatively small. We therefore expect the results of our work to not vary much if the proton beam energy is 80 GeV instead of 120 GeV.} striking a graphite target to generate the high-intensity neutrino beam. Of interest for this work, the proton collisions produce various hadronic final states such as pions, Kaons, $\eta$ mesons, etc. Specifically, we are focused on neutral and charged pions, Kaons and $\eta$ mesons. In their decays, these can produce the new BSM particles $Z'$ and $\varphi$. The proton beam is expected to deliver at least $1.47 \times 10^{21}$ protons on target (POT) per year, with potential upgrades throughout the lifecycle of the experiment. We will assume a nominal exposure of $1.47 \times 10^{22}$ POT for our analyses, conservatively corresponding to \textit{at most} ten years of data collection.

The DUNE near detector complex~\cite{AbedAbud:2021hpb}, including several detector components, is located 574 meters away from the target and therefore when the decays of the new particles $Z'$ and $\varphi$ occur far away from the production vertex, the final states can give a signal in the DUNE near detector. This setup is ideal in probing the parameter space in which $g_{BL}$ and the new particle masses are small, and therefore they are long-lived.

As detailed in Section~\ref{subsubsec:ZPrimeProduction}, in our $U(1)_{B-L}$  model, $\pi^0, \eta\to \gamma+Z'$ provides the source for the new BSM particle $Z'$. This decay will take place promptly, effectively in the DUNE target, and will be our source for $Z'$. There can also be direct bremsstrahlung  production of $Z'$ from $pp\to pp+Z'$ since protons have nonzero $B-L$ quantum number. The $\varphi$ production arises mostly from the decays $K^+\to \pi^++\varphi$ and $K_L \to \pi^0 + \varphi$.
Whether we are looking for $\varphi$ or $Z'$ signatures in the near detector, we will be interested in final states including opposite-sign pairs of charged particles. This includes $e^+ e^-$, $\mu^+\mu^-$, and pionic final states. The DUNE capability, specifically using its gaseous argon near detector (immediately downstream of the liquid argon near detector) to identify final-states of particles decaying in flight has been detailed in Refs.~\cite{Ballett:2019bgd,Berryman:2019dme,Coloma:2020lgy,Kelly:2020dda,Brdar:2020dpr,Dev:2021ofc, Breitbach:2021gvv}. We will use the results of these works and consider that a background-free search for these final states is possible.

As discussed in Section~\ref{sec:Introduction}, our search strategy naturally divides into three different categories based on the (non)existence of either the scalar $\varphi$ and gauge boson $Z'$. Phenomenologically speaking, we are interested in the four free parameters in Eq.~(\ref{eqn:parameters}). We divide the search strategies by considering the following cases:
\begin{itemize}
    \item \textbf{Pure \texorpdfstring{$Z'$}{Zp} case:} $\sin\vartheta = 0$ -- if a scalar boson exists at all, it is decoupled and does not have any impact on the phenomenology of $Z'$. This is explored in Section~\ref{subsec:PureZPrime}.

    \item \textbf{Pure \texorpdfstring{$\varphi$}{phi} case:} $g_{BL} = 0$ -- if a gauge boson exists at all, it is decoupled and does not impact the phenomenology of $\varphi$. This has been studied in Ref.~\cite{Berryman:2019dme}, and will not be detailed in this paper any more.

    \item \textbf{Combined \texorpdfstring{$Z'$}{Zp} and \texorpdfstring{$\varphi$}{phi} case (effects of $\varphi \to Z'Z'$):} If both particles are relevant at DUNE and $m_\varphi > 2m_{Z'}$ then the decay channel $\varphi \to Z' Z'$ can impact the prospects of the DUNE search for $\varphi$ -- we investigate this effect for DUNE and other $\varphi$ searches in Section~\ref{subsec:EffectonScalarLims}.
    Furthermore, this can also provide additional flux of $Z'$ at the DUNE near detector. This additional $Z'$ flux allows for increased sensitivity as a function of $m_{Z'}$ and $g_{BL}$, subject to current constraints on $m_\varphi$ and $\sin\vartheta$. This is discussed in Section~\ref{subsec:CombinedSearch} and is the main emphasis of this paper. As mentioned in the introduction, the associated production of both $Z'$ and $\varphi$ in some decay channels will be highly suppressed either by $g_{BL}^4$ or $g_{BL}^2 \sin^2\vartheta$, and can be neglected.
\end{itemize}

\subsection{Pure Gauge Boson Search}
\label{subsec:PureZPrime}

Here we discuss the case in which the only new physics particle is the gauge boson $Z'$ with a mass $m_{Z'}$ and gauge coupling $g_{BL}$. In the context of a search in a proton beam-dump environment, this scenario is very similar to the study performed for a $U(1)$ dark photon in Ref.~\cite{Berryman:2019dme}, with a mapping of the kinetic mixing parameter $\varepsilon$ onto this gauge coupling. This remapping is discussed for proton beam-dump experiments, as well as other experimental contexts, in Ref.~\cite{Bauer:2018onh}. In the following subsections, we discuss $Z'$ production in Section~\ref{subsubsec:ZPrimeProduction} and DUNE sensitivity in Section~\ref{subsubsec:ZPrimeSensitivity}; expressions for the decay of $Z'$ can be found in Section~\ref{sec:Model}.

\subsubsection{Production of Gauge Bosons}
\label{subsubsec:ZPrimeProduction}

The new gauge boson $Z'$ can be produced in the same ways as the $U(1)$ dark photon -- decays of neutral pseudoscalar mesons $\mathfrak{m} \to \gamma Z'$, and proton bremsstrahlung $p p \to p p Z'$. We give expressions for the flux of these $Z'$ at the DUNE near detector here.

The neutral pseudoscalar meson production flux is determined using
\begin{align}
    \Phi_{\mathfrak{m}Z'} &= \frac{c_\mathfrak{m} N_{\rm POT}}{A_{\rm Det.}} \epsilon(m_{Z'}) \mathrm{BR}\left(\mathfrak{m} \to \gamma Z'\right), \nonumber \\
    &= \frac{c_\mathfrak{m} N_{\rm POT}}{A_{\rm Det.}} \epsilon(m_{Z'}) \left[ 2\left(\frac{g_{BL}^2}{e^2}\right) \mathrm{BR}\left(\mathfrak{m} \to \gamma \gamma\right) \left( 1 - \frac{m_{Z'}^2}{m_\mathfrak{m}^2} \right)^3 \right].\label{eq:FluxZprimeM}
\end{align}
The quantities in Eq.~\eqref{eq:FluxZprimeM} are as follows: $c_\mathfrak{m}$ is the average number of mesons $\mathfrak{m}$ produced in a given POT collision; $N_{\rm POT}$ is the POT number considered in the experimental exposure; $A_{\rm Det.}$ is the detector area as viewed by an incoming particle; $\epsilon(m_{Z'})$ is the geometrical acceptance factor of the $Z'$ particles emerging from the decay $\mathfrak{m} \to \gamma Z'$, determined using Monte Carlo simulations; $e$ is the electric charge, $m_{\mathfrak{m}}$ the mass of $\mathfrak{m}$, and $\mathrm{BR}\left(\mathfrak{m} \to \gamma\gamma\right)$ is the SM BR of the meson $\mathfrak{m}$ into two photons. We find that the dominant contribution to $\Phi_{\mathfrak{m}Z'}$ comes from $\pi^0$ and $\eta$ mesons.

For proton-proton bremsstrahlung, we use the calculations of Refs.~\cite{Blumlein:2013cua,deNiverville:2016rqh}
 with specifics for DUNE discussed in Ref.~\cite{Berryman:2019dme}. The total flux from this bremsstrahlung process is
 \begin{equation}
     \Phi_{\mathrm{Brem}Z'} = \frac{N_{\rm POT}}{A_{\rm Det.} \sigma_{pN}(s)} \left\lvert F_{1,N}(m_{Z'}^2)\right\rvert^2 \int \mathrm{d}z\:  \int_{\rm Det.} \mathrm{d}p_T^2\: \sigma_{pN}\left(2m_p(E_p - E_{Z'})\right) w_{ba}(z,p_T^2) \,.
     \label{eq:ZpBrem}
 \end{equation}
 Here, $\sigma_{pN}$ is the total proton-target cross section evaluated at $s = 2E_p m_p$ with $E_p = 120$ GeV being the DUNE beam energy. $F_{1,N}$ is a form factor allowing for mixing between $Z'$ and the SM vector mesons, specifically when $m_{Z'}$ approaches $m_\rho$. The two integrals are performed over the variables $p_T^2$ (the transverse momentum squared of the outgoing $Z'$) and $z$, the fraction of the incoming proton's initial momentum that is transferred to the longitudinal momentum of the outgoing $Z'$, where we label ``Det.'' to indicate that the range of integration of these two is such that the outgoing $Z'$ is pointing from its production point to the front face of the near detector. The photon splitting function $w_{ba}$ is given in Refs.~\cite{Blumlein:2013cua,deNiverville:2016rqh,Berryman:2019dme} and, for the case of a $U(1)_{B-L}$ gauge boson, is proportional to $g_{BL}^2/4\pi$.

Other production mechanisms for $Z'$ are possible, however we find that all of them are subdominant to the neutral pseudoscalar meson decays and proton-proton bremsstrahlung. These include $K_L \to \gamma Z'$, three-body decays of charged mesons $\pi^+,\, K^+  \to \ell^+ \nu Z'$ (with $\ell = e,\,\mu$), loop-level flavor-changing decays such as $K \to \pi Z'$ (including both neutral and charged Kaon decays), and heavy baryon decay $\Delta \to N Z'$. For completeness, we give the BR expressions in these channels in Appendix~\ref{appendix:ZPrimeFluxes}.

\begin{figure}[!t]
  \centering
  \includegraphics[width=0.7\textwidth]{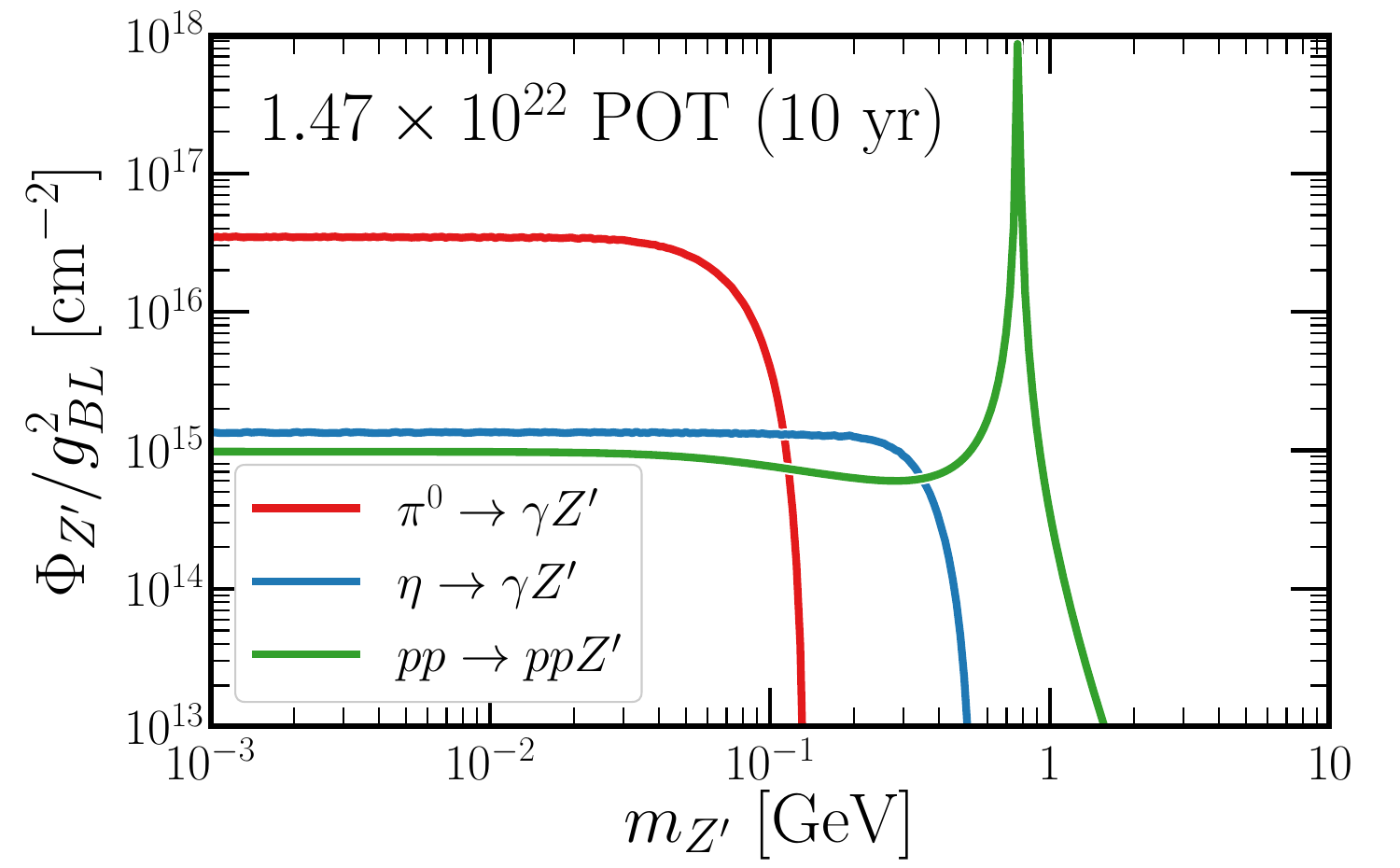}
  \caption{Expected flux of $Z'$ gauge bosons at the DUNE near detector in the channels of $\pi^0 \to \gamma Z'$, $\eta \to \gamma Z'$ and $pp \to pp Z'$, assuming the particles are all infinitely long-lived (and have not decayed en route). This flux scales with $g_{BL}^2$ and has been determined assuming $1.47 \times 10^{22}$ protons striking the DUNE target, a conservative estimate for ten years of operation of the experiment.
  }
  \label{fig:flux:Zprime}
\end{figure}

As discussed earlier in this Section, we will assume ten years of data collection.\footnote{Planned upgrades to the PIP-II beam will increase this rate. Our sensitivity can be viewed then as what can be done with \textit{at most} ten years of data collection.} Assuming only $Z'$ exists (either $\varphi$ does not exist, $\sin\vartheta \to 0$, or $m_\varphi < 2m_{Z'}$), we can determine the flux of $Z'$ at the DUNE Near Detector complex using the above. This is shown, divided by different contributions, in Fig.~\ref{fig:flux:Zprime}.

Given the flux of $Z'$, we may determine the number of decay events in the DUNE Near Detector volume using
\begin{equation}
	\label{eq:DecaySigEvts}
    N_{\rm Sig.} = \int_{E_{\rm min.}}^{E_{\rm max.}} \mathrm{d}E_{Z'} \:
    \int_{A_{\rm Det.}} \mathrm{d}A\:
    \int_{D_{\rm Det.}}^{D_{\rm Det.} + L_{\rm Det.}}  \mathrm{d}x \: \left[\frac{{\rm d}\Phi_{Z'}}{{\rm d} E_{Z'}} P_{\rm Decay} (E_{Z'}, x)\right] .
\end{equation}
The outermost integral is over the energy of the $Z'$ bosons and can account for detector thresholds, efficiencies, etc. (with $E_{\rm min}$ and $E_{\rm max}$ respectively the minimal and maximal energy of $Z'$ bosons) -- in our simulation we allow for the full range of $E_{Z'}$ and assume 100\% signal identification efficiency. The second integral is over the surface area of the detector, $2.5$ m in radius for the gaseous argon time projection chamber. The innermost integral is over the depth of the detector $L_{\rm Det.} = 5$ m with $D_{\rm Det.}$ being the distance to the detector, $579$ m\footnote{We consider only decays in the gaseous argon detector, situated directly behind the liquid argon one with a depth of $5$ m.}. The probability of a decay occurring at position $x$ is
\begin{equation}
    P_{\rm Decay}(E_{Z'}, x) = \frac{1}{\gamma_{Z'} c\tau_{Z'}} e^{-\frac{x}{\gamma_{Z'} c\tau_{Z'}}}.
\end{equation}
In this decay expression, $\gamma_{Z'} = E_{Z'}/m_{Z'}$ is the Lorentz boost factor and $\tau_{Z'}$ is the proper lifetime of $Z'$, which can be obtained using the expressions for its width given in Section~\ref{sec:Model}.

The form of Eq.~\eqref{eq:DecaySigEvts} assumes that the entire $Z'$ flux $\mathrm{d}\Phi_{Z'}/\mathrm{d}E_{Z'}$ originates at a common location, assumed to be at the proton/target interaction location. If the $Z'$ flux is produced continuously along the beam pipe, Eq.~\eqref{eq:DecaySigEvts} will require an additional integral over the $Z'$ production distribution. This will be relevant when we consider $Z'$ coming from $\varphi \to Z' Z'$ decay in Section~\ref{subsec:CombinedSearch}. With Eq.~\eqref{eq:DecaySigEvts}, we now can obtain the sensitivity of DUNE to the pure-$Z'$ scenario.

\subsubsection{Sensitivity to Pure Gauge Boson Scenario}\label{subsubsec:ZPrimeSensitivity}
\begin{figure}[!t]
  \centering
  \includegraphics[width=0.7\textwidth]{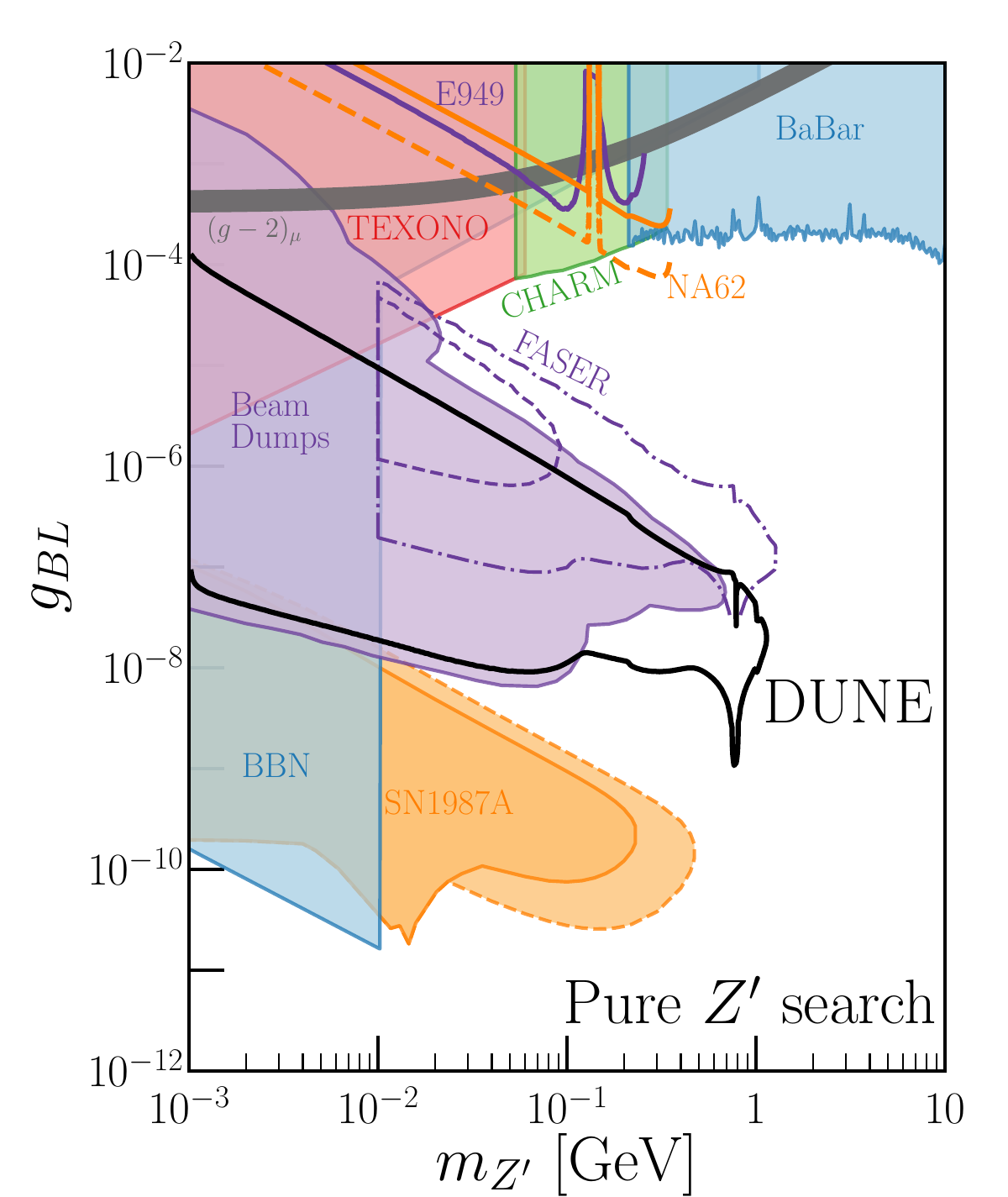}
  \caption{Prospects of the Pure $Z'$ search without the scalar $\varphi$ using the DUNE Near Detector (black) where the black lines encompass parameter space for which at least ten signal events (zero background assumed) are expected.
  Current limits from beam-dump experiments (shaded purple) from Ref.~\cite{Bauer:2018onh}, TEXONO (shaded red)~\cite{Deniz:2009mu, Bilmis:2015lja}, CHARM (shaded green)~\cite{Bergsma:1985qz}, BaBar (shaded cyan)~\cite{Lees:2014xha}, E949 (purple line)~\cite{Artamonov:2009sz} and NA62 (orange line)~\cite{Ruggiero:2019}, and BBN (shaded blue)~\cite{Knapen:2017xzo} are also shown. The dark (solid) and faint (dashed) orange regions are respectively the conservative and aggressive SN1987A limits~\cite{Croon:2020lrf}. The gray band denotes the preferred region for muon $g-2$ anomaly at the $2\sigma$ C.L.~\cite{Bennett:2006fi, Abi:2021gix}. Also shown are the prospects from future NA62 data (dashed orange line)~\cite{Anelli:2005ju}, FASER (dashed purple line) and FASER2 (dot-dashed purple line)~\cite{Ariga:2018uku}. See text for more details.
  \label{fig:prospect:PureZprime}}
\end{figure}

Here, we detail the DUNE sensitivity to the pure-$Z'$ scenario, in which the scalar $\varphi$ either does not exist or its decay into $Z'$ pairs is rare. The sensitivity is depicted in Fig.~\ref{fig:prospect:PureZprime}, where the solid black line demonstrates the DUNE discovery capability: in the region within the black line, at least ten signal events of $Z' \to \ell^+ \ell^-$ are expected in ten years of data collection. As argued in Ref.~\cite{Berryman:2019dme}, this constitutes a statistically significant signal, especially since the invariant mass of $m_{Z'}$ can be reconstructed by the fully visible decay. The DUNE sensitivity in the visible channel dies off for $m_{Z'}<2m_e$, because the $Z'$ in that case dominantly decays to invisible final states of light neutrino pairs.\footnote{The additional neutrino flux from $Z'\to \nu\bar{\nu}$ decay can in principle be used to extend the light $Z'$ reach at DUNE below the electron threshold, studied in the context of neutrinophilic vector bosons in Ref.~\cite{Bakhti:2018avv}. Other complementary probes include the neutrino trident production and elastic neutrino-electron  scattering~\cite{Ballett:2019xoj,Altmannshofer:2019zhy}}

In the pure-$Z'$ scenario, we see that DUNE can extend on existing beam-dump searches (shown as a purple region, collected from Ref.~\cite{Bauer:2018onh}). This includes the E141~\cite{Riordan:1987aw,Bjorken:1988as,Bjorken:2009mm,Andreas:2012mt}, Orsay~\cite{Davier:1989wz}, NuCal~\cite{Blumlein:2011mv,Blumlein:2013cua,Tsai:2019mtm}, E137~\cite{Bjorken:1988as}, and LSND~\cite{Athanassopoulos:1997er} experiments. In contrast, DUNE can reach heavier $m_{Z'}$ (due to large $\eta$ meson production and bremsstrahlung contributions) and lower $g_{BL}$ (due to the large target-detector distance). We also show the projected sensitivity of the FASER experiment near the LHC in its first run (dashed purple) and proposed FASER2 run (dot-dashed purple)~\cite{Ariga:2018uku}. FASER is sensitive to similar $m_{Z'}$ with larger $g_{BL}$, since the $Z'$ produced near the LHC collision point have larger boosts (and therefore longer lab-frame lifetime) than in the DUNE environment. In this sense, DUNE as a detector searching for decaying new-physics particles is highly complementary to FASER. In the time between now and DUNE data collection, other Fermilab-based neutrino experiments could reach into this parameter space as well, including the Short-Baseline Neutrino facility detectors. See, e.g., Ref.~\cite{Batell:2019nwo} for a study on searches for dark scalars in this setup.

For relatively larger gauge coupling and $Z'$ masses, there are limits on $m_{Z'}$ and $g_{BL}$ from the proton beam-dump experiment CHARM~\cite{Bergsma:1985qz}, the searches of $Z'$ in the channel $e^+ e^- \to \gamma Z' \to \gamma \ell^+ \ell^-$ (with $\ell = e,\;\mu$) at BaBar~\cite{Lees:2014xha}, and the neutrino scattering experiment TEXONO~\cite{Deniz:2009mu, Bilmis:2015lja}, which are shown respectively as the green, cyan and red shaded regions in Fig.~\ref{fig:prospect:PureZprime}. Belle II can improve on the BaBar limits~\cite{Abe:2010gxa}, extending to ever lower $g_{BL}$ as it collects data~\cite{Kou:2018nap}. More limits can be found in Refs.~\cite{Aubert:2009cp, Bauer:2018onh}. For completeness, we also present our calculated limits on $K^+ \to \pi^+ Z' \to \pi \nu\bar\nu$ from E949~\cite{Artamonov:2009sz} and NA62~\cite{Ruggiero:2019}, which are however weaker than the limits from TEXONO, CHARM and NA62, as indicated by the purple and orange lines in Fig.~\ref{fig:prospect:PureZprime}. The future NA62 data can improve the measurement of ${\rm BR} (K^+ \to \pi^+ \nu\bar\nu)$ by roughly one order of magnitude~\cite{Anelli:2005ju}, thus extend on the existing limits from CHARM and BaBar, as denoted by the dashed orange line in Fig.~\ref{fig:prospect:PureZprime}. More details of the E949 and NA62 limits and the prospects at future NA62 can be found in Appendix~\ref{appendix:Zprimelimit}.

We note in passing that although a light $Z'$ boson of $U(1)_{B-L}$ generates a positive contribution to the muon anomalous magnetic moment~\cite{Lindner:2016bgg}, the $2\sigma$ parameter space preferred by the $(g-2)_\mu$ anomaly~\cite{Bennett:2006fi, Abi:2021gix} is deep inside the exclusion region in Fig.~\ref{fig:prospect:PureZprime}, as shown by the gray band. Thus, if the $(g-2)_\mu$ anomaly turns out to be a true signal of BSM physics\footnote{This is subject to further scrutiny, given that a recent lattice result~\cite{Borsanyi:2020mff} for the SM prediction is much closer to the experimental value, as compared to the lattice world average evaluated in Ref.~\cite{Aoyama:2020ynm}.}, the minimal $U(1)_{B-L}$ model with a flavor-universal $Z'$ coupling needs to be extended to explain this result; see, e.g., Refs.~\cite{Ma:2001md, Altmannshofer:2016oaq, Altmannshofer:2016brv,CarcamoHernandez:2019ydc, Dev:2020drf,Abdallah:2020biq,Bodas:2021fsy,Amaral:2021rzw, Athron:2021iuf}.

Finally, at low masses and small couplings, constraints from Big Bang Nucleosynthesis (BBN) and the observation of Supernova 1987A (SN1987A) apply, which are shown respectively as the blue and orange shaded regions in Fig.~\ref{fig:prospect:PureZprime}. In the early universe, the $Z'$ boson may decay into neutrinos and electrons (if kinematically allowed) and thus contribute to the evolution of neutrinos and photons at the MeV scale. This could spoil the success of BBN and thus is constrained by the precise measurement of the light degrees of freedom $N_{\rm eff}$ by Planck~\cite{Aghanim:2018eyx}. However, the $N_{\rm eff}$ limits on the $B-L$ $Z'$ boson depend largely on the $Z'$ mass and the coupling $g_{BL}$, which dictates whether the $Z'$ boson could reach equilibrium with the SM particles at the MeV scale. Moreover, neutrino oscillations are also important for the $N_{\rm eff}$ limits~\cite{Escudero:2018mvt, Escudero:2020dfa}. The dedicated calculations of the cosmological limits are beyond the main scope of this paper. For simplicity, we take the BBN limit from Ref.~\cite{Knapen:2017xzo}. Other relevant studies can be found, e.g., in Refs.~\cite{Heeck:2014zfa, Kamada:2018zxi, Escudero:2019gzq, Dutta:2020jsy}. The supernova constraints (dark orange region/solid lines corresponding to conservative constraints, and faint orange region/dashed lines corresponding to less conservative) come from Ref.~\cite{Croon:2020lrf} -- we direct the reader to this reference, as well as Ref.~\cite{Knapen:2017xzo,Chang:2016ntp,Rrapaj:2015wgs} for more discussion on supernova luminosity constraints on $U(1)_{B-L}$ vector bosons and associated models. We conclude this discussion by noting that both the BBN and SN1987A limits are model dependent -- for example, the chameleon effect (due to the environmental matter density) can weaken astrophysical limits~\cite{Nelson:2008tn}, and late reheating can open up parameter space with respect to BBN constraints~\cite{Depta:2020wmr}. This model-dependence will be more relevant when we revisit this parameter space in Section~\ref{subsec:CombinedSearch} where we perform a combined search for $Z'$ and the scalar boson $\varphi$ -- the interplay of these two new particles and their associated mass scales complicates the calculations of the references mentioned above. We leave a detailed analysis of astrophysical constraints on the combined $Z'$, $\varphi$ scenario to future work.

\subsection{Effects of Gauge Coupling on Searches for Scalar \texorpdfstring{$\varphi$}{phi}}
\label{subsec:PureScalar}

Now we shift focus to the prospects of detecting the scalar boson $\varphi$ responsible for breaking the $U(1)_{B-L}$ symmetry. Its couplings to SM particles are controlled by the mixing angle which we refer to as $\sin\vartheta$. Section~\ref{sec:Model} detailed the different decay channels and widths of $\varphi$. In Section~\ref{subsubsec:ScalarProduction} we
discuss the production of $\varphi$ in the DUNE environment. Section~\ref{subsec:EffectonScalarLims} then explores how the existence of both $\varphi$ and $Z'$ (and the decay channel $\varphi \to Z' Z'$) can hinder current and future searches for the scalar boson $\varphi$. The improvement of $Z'$ prospects at DUNE as a result of the $\varphi\to Z' Z'$ source will be investigated in Section~\ref{subsec:CombinedSearch}.

\subsubsection{Production of Scalar Boson}\label{subsubsec:ScalarProduction}
From mixing with the SM Higgs, the light scalar $\varphi$ can be produced in/near the DUNE target in the following channels:
\begin{itemize}
  \item Loop-level flavor-changing meson decays $K^\pm \to \pi^\pm \varphi$ and $K_L \to \pi^0 \varphi$. This is identical to the production in Fig.~6.1 of Ref.~\cite{Berryman:2019dme}, with the widths~\cite{Dev:2019hho}
  \begin{eqnarray}
\label{eqn:Kdecay:U1:1}
\Gamma (K^\pm \to \pi^\pm \varphi) &\ \simeq \ &
\frac{m_{K^\pm} \left| y_{sd} \right|^2 \sin^2\vartheta}{64 \pi}
\frac{m_{K^\pm}^2}{m_\varphi^2} \lambda (m_{K^\pm}, m_{\pi^\pm}, m_{\varphi}) \,, \\
\label{eqn:Kdecay:U1:2}
\Gamma (K_L \to \pi^0 \varphi) &\ \simeq \ &
\frac{m_{K_L} \left( {\rm Re} \, y_{sd} \right)^2 \sin^2\vartheta}{64 \pi}
\frac{m_{K^0}^2}{m_\varphi^2} \lambda (m_{K_L}, m_{\pi^0}, m_{\varphi}) \,,
\end{eqnarray}
  where the loop-level FCNC coupling~\cite{Batell:2009jf}
  \begin{eqnarray}
  y_{sd} \ = \ \frac{3\sqrt2 G_F m_t^2 V_{ts}^\ast V_{td}}{16\pi^2} \,
\frac{m_\varphi}{\sqrt2 v_{\rm EW}} \, ,
\label{eqn:ysd}
\end{eqnarray}
  which is induced by the $W-$top loop, and $v_{\rm EW} = (\sqrt2 G_F)^{-1/2}/\sqrt2 \simeq 174$ GeV is the electroweak VEV. Note that the partial decay widths in Eqs.~(\ref{eqn:Kdecay:U1:1}) and (\ref{eqn:Kdecay:U1:2}) are almost identical, except for the crucial difference that the decay $K_L \to \pi^0 \varphi$ depends only on the real part of the coupling $y_{sd}$.

  \item The bremsstrahlung process $pp \to pp \varphi$ is analogous to the $pp\to pp Z'$ process discussed in Eq.~\eqref{eq:ZpBrem}. Ref.~\cite{Foroughi-Abari:2020gju} found that this production process is dominant for such a search at LSND (proton beam energy of 800 MeV). In contrast to the $Z'$ bremsstrahlung process, the scalar coupling is suppressed by the effective proton Yukawa coupling. Ref.~\cite{Boiarska:2019jym} found that this coupling is small, $\mathcal{O}(10^{-3})$, and Ref.~\cite{Batell:2019nwo} used this result to find that the proton bremsstrahlung process is highly suppressed relative to the Kaon decay processes discussed above for scalar production in the NuMI beam environment (proton beam energy of 120 GeV). We have verified that it is also highly suppressed in the DUNE environment.
\end{itemize}

In Eq.~\eqref{eq:DecaySigEvts} we provided the expected number of decay signal events using some flux of (potentially) decaying particles. The same formalism applies here with the $\varphi$ flux instead of $Z'$, however, as alluded to previously, an extra integral must be applied. Because the Kaons $K^\pm$ and $K_L$ are long-lived, the production of $\varphi$ is \textit{not} all in/near the DUNE target -- the distribution of $\varphi$ production points spans hundreds of meters\footnote{The DUNE decay pipe ends roughly 230 meters from the target, so the $\varphi$ can be produced over this entire distance, including some component from Kaon decay-at-rest at the absorber at the end of the decay pipe.}.

\subsubsection{Effects on \texorpdfstring{$\varphi$}{phi} prospects at DUNE}
\label{subsec:EffectonScalarLims}

\begin{figure}[!t]
\begin{center}
\includegraphics[width=0.78\linewidth]{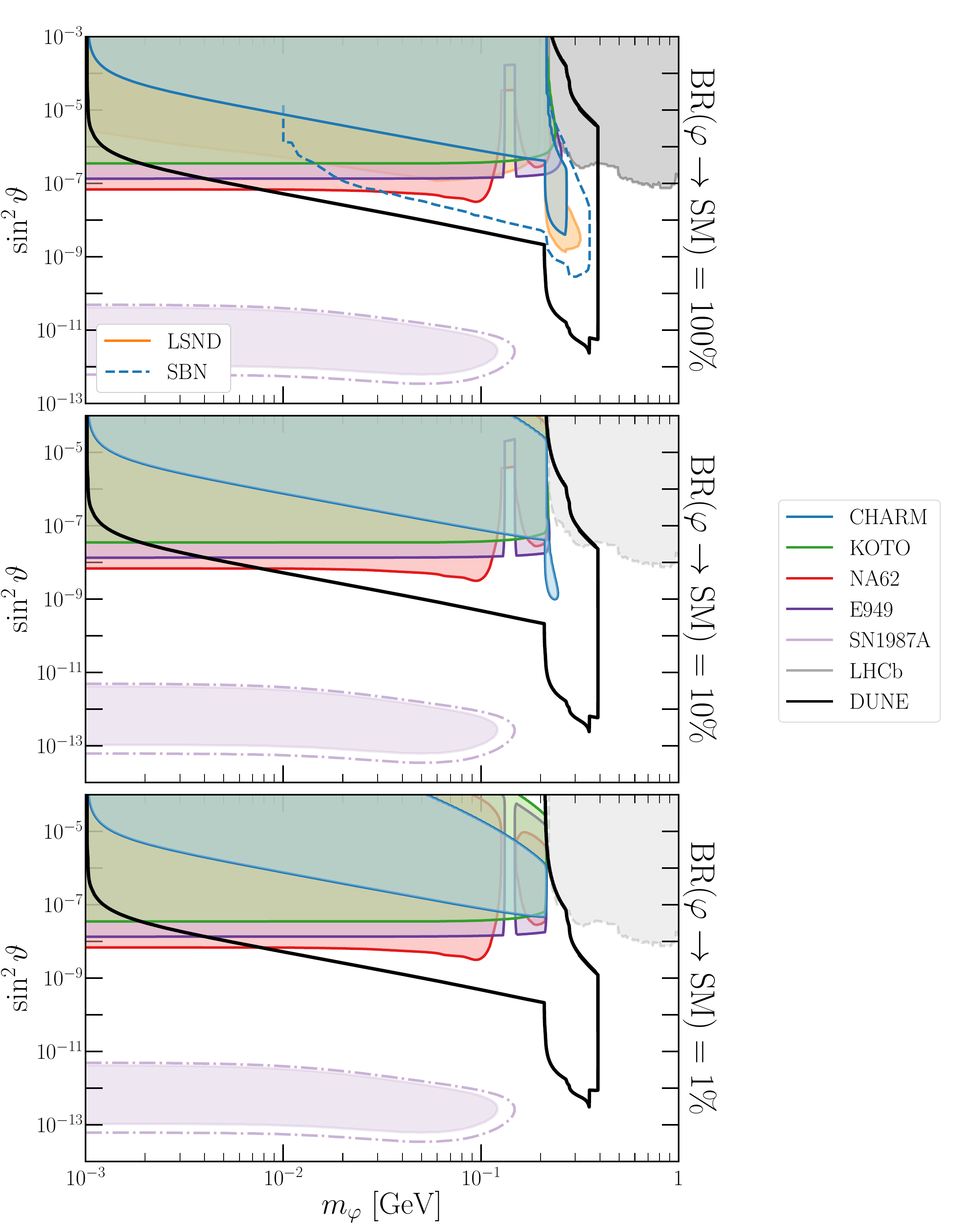}
\caption{Limits on the scalar mass $m_\varphi$ and the mixing parameter $\sin^2\vartheta$. For each panel, we make a different assumption about the BR of $\varphi$ into SM particles. The left panel assumes 100\%, the center 10\%, and the right 1\%. We include constraints from CHARM (blue)~\cite{Bergsma:1985qz,Winkler:2018qyg}, KOTO (green)~\cite{Ahn:2018mvc}, NA62 (red)~\cite{Ruggiero:2019}, E949 (purple)~\cite{Artamonov:2009sz}, LHCb (gray)~\cite{Aaij:2015tna,Aaij:2016qsm}, projections from DUNE (black) in all panels, as well as constraints from LSND (orange)~\cite{Foroughi-Abari:2020gju} and SBN projections  from SBND and ICARUS (dashed blue)~\cite{Batell:2019nwo} in the left panel. The supernova constraints assuming the luminosities of $3\times 10^{53}$ erg/sec and $5\times 10^{53}$ erg/sec are shown respectively by the dot-dashed purple contours and purple shaded regions~\cite{Dev:2020eam,merger}.
\label{fig:scalar:varyBr}}
\end{center}
\end{figure}

The decay $\varphi \to Z'Z'$ not only opens up a new decay channel, but also contributes to the total width of $\varphi$. Therefore, this extra channel will affect any existing search for $\varphi$ by (a) modifying its branching fraction into visible final-state particles and (b) shortening its lifetime. In general, the constraints on $\varphi$ from searches for decays like $K^+ \to \pi^+ \nu \bar{\nu}$ (which could be mimicked by $K^+ \to \pi^+ \varphi$ and $\varphi$ decaying invisibly/outside a detector) and beam-dump searches like that of DUNE can be understood by two pieces. At the bottom of these exclusion regions, the reach of an experiment is limited by exposure (and/or backgrounds) -- for lower values of $\sin^2\vartheta$, not enough $\varphi$ can be produced to create a signal. At the top of these exclusion regions, the lifetime of $\varphi$ is so short that it decays away before leaving an imprint in a detector. This is particularly relevant for beam-dump searches like CHARM, where the target (production point of $\varphi$) and detector are well separated.

We reproduce existing limits on $m_\varphi$ and $\sin^2\vartheta$ coming from a variety of experiments -- CHARM~\cite{Bergsma:1985qz,Winkler:2018qyg}, KOTO~\cite{Ahn:2018mvc}, NA62~\cite{Ruggiero:2019}, and E949~\cite{Artamonov:2009sz} -- in light of this modification. We also reproduce the DUNE sensitivity from Ref.~\cite{Berryman:2019dme}, including this new decay channel. We do so by allowing the branching fraction of $\varphi$ into SM particles, denoted as BR($\varphi \to\ \mathrm{SM}$), to vary as a free parameter. In actuality, for a given ($m_\varphi,~\sin^2\vartheta$), this BR is dependent on those parameters as well as $m_{Z'}$ and $g_{BL}$. The limits from these experiments, including the modifications, are shown in Fig.~\ref{fig:scalar:varyBr} for BR($\varphi \to\ \mathrm{SM}$) = 100\% (top), BR($\varphi \to\ \mathrm{SM}$) = 10\% (center), and BR($\varphi \to\ \mathrm{SM}$) = 1\% (bottom). Changes to the no-$Z'$ case (effectively, the left panel), are most apparent at higher $m_\varphi$ and larger $\sin^2\vartheta$, e.g. the ``lobe'' in the CHARM exclusion closes up.

In Fig.~\ref{fig:scalar:varyBr}, we also include several other results -- in gray, we include constraints from rare $B$ meson decays at LHCb~\cite{Aaij:2015tna,Aaij:2016qsm}. We do not perform our own simulation of how these limits change for BR($\varphi\to\ \mathrm{SM}$) $\neq 1$, however we expect that their behavior in the mass/angle range of interest will not vary much. In our simulations in Section~\ref{subsec:CombinedSearch}, we do not allow $\sin^2\vartheta > 10^{-6}$ for any $m_\varphi$, in addition to requiring it to be consistent with all the experimental limits shown in Fig.~\ref{fig:scalar:varyBr}. Also, in the top panel of Fig.~\ref{fig:scalar:varyBr}, we include a limit from a reinterpretation of the LSND experiment's results~\cite{Foroughi-Abari:2020gju} and future projections of the Fermilab SBN detectors, a combination of using the SBND and ICARUS experiments with the BNB and NuMI beams as proposed in Ref.~\cite{Batell:2019nwo}. Due to the complicated nature of simulating the LSND and SBN setups, we do not provide these limits/projections in the center/bottom panels, where they should in principle exist as well.

As with the light gauge boson providing a positive contribution to the muon anomalous magnetic moment, so does a light neutral scalar~\cite{Lindner:2016bgg}. However, it requires $\sin^2\vartheta \simeq 1$ mixing with Higgs for our light scalar $\varphi$ to explain the  $(g-2)_\mu$ anomaly~\cite{Abi:2021gix}, which is certainly excluded in  Fig.~\ref{fig:scalar:varyBr} (well above the range we present). In general, a light scalar explanation of the $(g-2)_\mu$ anomaly must involve  lepton-flavor-violating couplings to be consistent with the existing constraints; see, e.g.~Refs.~\cite{Dev:2017ftk,Cherchiglia:2017uwv, Dev:2018upe, Chun:2019oix, Wang:2021fkn}.

BBN can also provide constraints on the existence of $\varphi$ if it is in thermal equilibrium and has yet to decay before the formation of light elements. If the lifetime $\tau_\varphi \gtrsim 1$ sec, the scalar $\varphi$ will contribute a factor of $4/7$ to the light degrees of freedom $N_{\rm eff}$~\cite{Kainulainen:2015sva, Fradette:2017sdd}. Although the scalar contribution to $N_{\rm eff}$ is smaller by a factor of three compared to $Z'$, it is still larger than the allowed range of $\Delta N_{\rm eff}$ by Planck at the time of CMB formation, which is roughly $0.2-0.5$ depending on the data sets adopted~\cite{Aghanim:2018eyx}. However, to implement the $\Delta N_{\rm eff}$ limits, $\varphi$ has to be in thermal equilibrium with the SM particles at the BBN temperature of MeV scale~\cite{Dev:2017dui}. The decaying of $\varphi$ into $e^+ e^-$ and $\gamma\gamma$ are both very small, therefore there is almost no parameter space of $m_{\varphi}$ and $\sin\theta$ to satisfy both the conditions above.

The supernova limits on the scalar $\varphi$ have been investigated in Refs.~\cite{Ishizuka:1989ts,Hanhart:2000er,Arndt:2002yg, Diener:2013xpa, Krnjaic:2015mbs, Lee:2018lcj,  Dev:2020eam}. In the supernova core, the scalar $\varphi$ can be produced via the bremsstrahlung process $N N \to N N \varphi$ (with $N$ being nucleons). Taking into account the partial cancellation between different Feynman diagrams~\cite{Dev:2020eam}, the observed neutrino luminosity ${\cal L}_\nu$ can be used to set limits on $m_\varphi$ and $\sin^2\vartheta$. Setting ${\cal L}_\nu = 3\times 10^{53}$ erg/sec and $5\times 10^{53}$ erg/sec respectively, the regions inside the dot-dashed light-purple contours and shaded regions in Fig.~\ref{fig:scalar:varyBr} can be excluded. The production amplitudes for the diagrams with the scalar $\varphi$ coupling to nucleons in Ref.~\cite{Dev:2020eam} have been corrected by a factor of $1/2$~\cite{merger}. One should note that in the supernova limits here we do not include the possible decay channel $\varphi \to Z'Z'$. If this channel is kinematically allowed, the supernova limits on $\varphi$ (and also $Z'$) could be potentially changed, but this requires a dedicated calculation beyond the scope of this paper, and is left for future work.

\subsection{Improving \texorpdfstring{$Z'$}{Zp} prospects at DUNE}
\label{subsec:CombinedSearch}

As discussed in Eq.~\eqref{eqn:phiZpZp}, if both $\varphi$ and $Z'$ exist and $m_\varphi > 2m_{Z'}$, then the decay $\varphi \to Z' Z'$ may occur. In Section~\ref{subsec:EffectonScalarLims} we discussed how this decay channel modifies limits on the scalar boson in the $(m_\varphi,~\sin^2\vartheta)$ plane. In this subsection, we will discuss how the additional flux of $Z'$ from this channel can improve search capabilities in the $(m_{Z'},~g_{BL})$ plane.

Before analyzing this flux contribution, we briefly note that additional associated production of \textit{both} $\varphi$ and $Z'$ may occur from some meson decays. In the presence of the gauge coupling of $\varphi$ to $Z'$ bosons, we can in principle have the associated production of $Z'$ and $\varphi$ from meson decays, for instance
\begin{eqnarray}
\pi^0 \to \gamma + Z' + \varphi\,, \quad
K \to \pi + Z' + \varphi \,, \quad
\rho \to Z' + \varphi \,.
\end{eqnarray}
If the scalar $\varphi$ is emitted from the $Z'$ boson line, the partial widths will be proportional to $g_{BL}^4$~\cite{Batell:2009di}. If the scalar $\varphi$ is emitted from the quark fermion lines, the widths will be proportional to $\sin^2\vartheta g_{BL}^2$. Therefore for small $\sin\vartheta$ and $g_{BL}$, these associated production channels are highly suppressed compared to the channels above. Therefore, we do not consider these channels in this paper.

Since the main source of $\varphi$ arises from decays of long-lived Kaons (compared to the prompt direct production of $Z'$ -- neutral, short-lived meson decays and proton-proton bremsstrahlung), the $Z'$ coming from $\varphi$ decays will originate all along the distance between the beam target and the end of the decay pipe. We take this into account by both allowing the Kaons to be long lived and the $\varphi$ to have a lifetime according to the relevant parameters -- $m_\varphi$, $\sin^2\vartheta$, $g_{BL}$, and $m_{Z'}$. With these ingredients, we can determine the resulting $Z'$ flux, as well as the number of $Z' \to e^+e^-$ and $Z' \to \mu^+ \mu^-$ events in our detector.

\begin{figure}[!ht]
  \centering
  \includegraphics[width=0.9\linewidth]{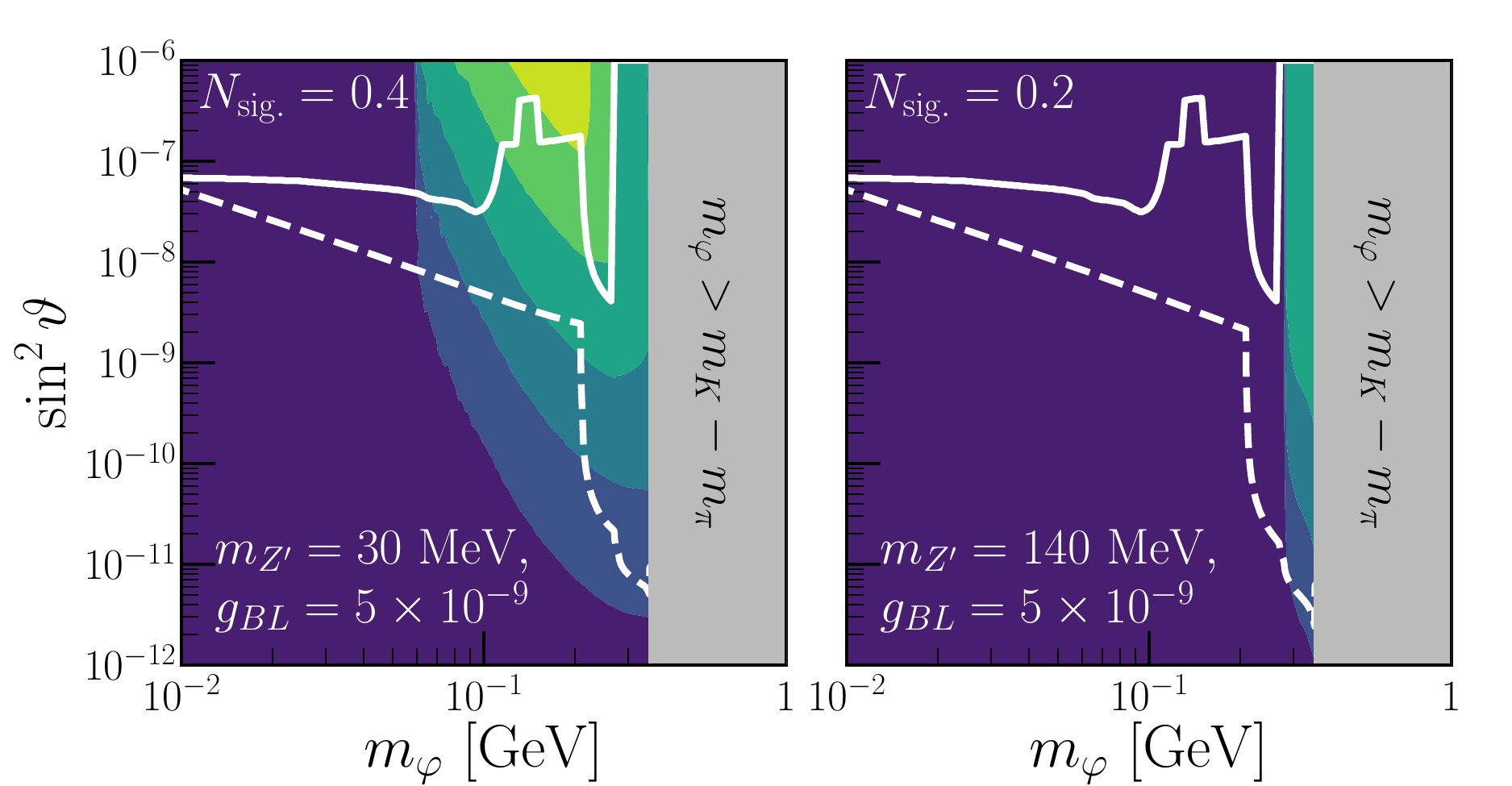}
  \caption{Expected number of signal events of $Z'$ decays in ten years of data collection at DUNE, assuming two benchmark parameters as given in Eqs.~\eqref{eqn:BP1} (left panel) and \eqref{eqn:BP2} (right panel). The different colored contours represent increasing signal events (from darker to lighter colors) by an order of magnitude each, purple corresponding to $N_{\rm sig.} < 1$.
  In both panels, when $m_\varphi < 2m_{Z'}$, the contribution is purely from direct production (meson decays and proton-proton bremsstrahlung), where $\varphi \to Z' Z'$ production increases the signal rate when $m_\varphi > 2m_{Z'}$. In white solid lines we show a collection of current constraints from Section~\ref{subsec:EffectonScalarLims} on $m_\varphi$ and $\sin^2\vartheta$, and in dashed white we show the expected DUNE sensitivity to search for the $\varphi$ decays. In the vertical gray shaded regions, the production of $\varphi$ from Kaon decay is kinematically forbidden.}
  \label{fig:CombinedExplanation}
\end{figure}

Fig.~\ref{fig:CombinedExplanation} demonstrates how we determine this. For a given combination of ($m_{Z'}$, $g_{BL}$) -- for example the BPs in Eqs.~(\ref{eqn:BP1}) and (\ref{eqn:BP2}) -- we determine the flux of $Z'$ at the DUNE Near Detector complex from ``direct'' production, i.e. the available neutral meson decays and proton-proton bremsstrahlung. That, combined with the $Z'$ lifetime (determined also by $m_{Z'}$ and $g_{BL}$) and BR into visible final states, determines the number of signal events we observe. For these two combinations, we would expect ${\sim}0.4$ or ${\sim}0.2$ events in ten years of data collection at DUNE - not enough to warrant a significant discovery. However, if there exists $\varphi$ with mass greater than 2$m_{Z'}$, then an additional $Z'$ flux can be generated. The total flux, and the number of signal events, would depend on the mass of $\varphi$ and the mixing $\sin^2\vartheta$, however, we cannot take $\sin^2\vartheta$ to be arbitrarily large, given existing constraints discussed in Section~\ref{subsec:EffectonScalarLims} and Fig.~\ref{fig:scalar:varyBr}. In the left panel of Fig.~\ref{fig:CombinedExplanation}, the colored contours display the number of expected signal events for the BP in Eq.~(\ref{eqn:BP1}) as we vary $m_\varphi$ and $\sin^2\vartheta$, each color change representing an order-of-magnitude increase in $N_{\rm sig.}$ from bottom to top (the purple regions correspond to $N_{\rm sig.} < 1$ and each subsequent color change goes to $1 \leq N_{\rm sig.} \leq 10$ and so on). We scan over this parameter space, determining the maximum $N_{\rm sig.}$ subject to the existing constraints in this space (solid white line) as well as the future DUNE sensitivity (dashed white line). For this combination of $m_{Z'}$ and $g_{BL}$, we obtain $N_{\rm sig.}$ as large as ${\sim}10^4$ in ten years of data collection (this occurs if $m_\varphi \approx 150$ MeV and for $\sin^2\vartheta \approx 3\times 10^{-7}$, where the colored contours in Fig.~\ref{fig:CombinedExplanation} left panel are lightest), which would be detectable in DUNE. In contrast, we show the result for the BP in Eq.~(\ref{eqn:BP2}) with a heavier $Z'$ in the right panel of Fig.~\ref{fig:CombinedExplanation}, where the color changing denotes also the one-order-of-magnitude higher from bottom to top. As a result of the heavier $Z'$ mass, there is only a narrow window for $m_\varphi$ with the enhanced $Z'$ production rate at the DUNE, i.e. $2m_{Z'} < m_\varphi < m_K - m_\pi$, or equivalently $280 \, {\rm MeV} < m_\varphi \lesssim 363$ MeV. Beyond the mass threshold of $m_K - m_\pi$, the scalar $\varphi$ can not be produced from the flavor-changing Kaon decays, which is shown as the vertical gray shaded regions in both panels of Fig.~\ref{fig:CombinedExplanation}.

\begin{figure}[!t]
  \centering
  \includegraphics[width=0.7\textwidth]{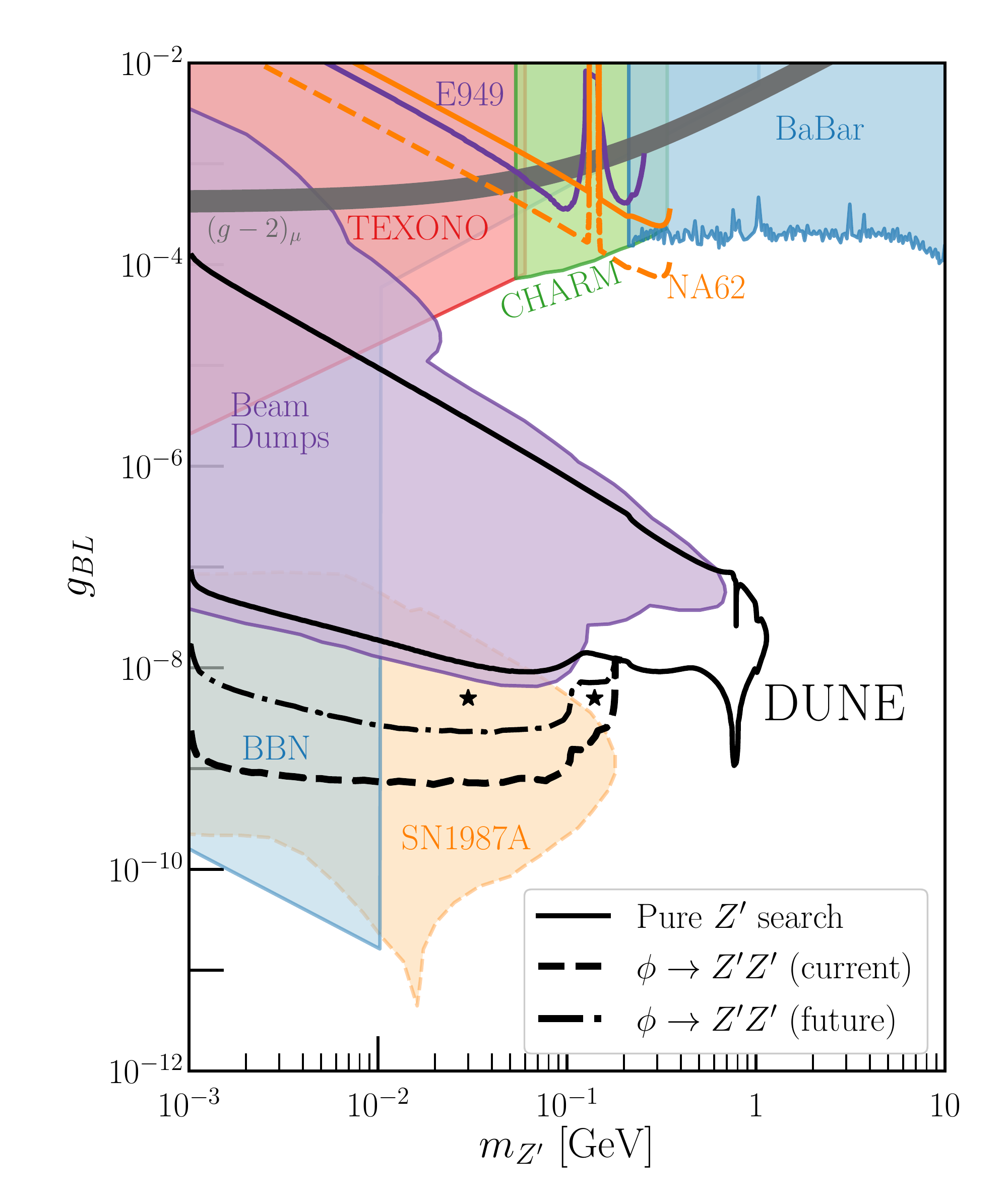}
  \caption{Improvements of the $Z'$ prospects at DUNE near detector complex as a result of the new channel $\varphi \to Z'Z'$. The dashed line indicates the improvement if we only subject $\varphi$ and $\sin^2\vartheta$ to current constraints from CHARM, KOTO, NA62, and E949 (see Section~\ref{subsec:EffectonScalarLims}), where the dot-dashed line also includes constraints from DUNE's direct search for $\varphi$. Between the dot-dashed and dashed lines, any $Z'$ signal in DUNE should also require a direct $\varphi$ signal in the same data. All other constraints and labels are the same as in Fig.~\ref{fig:prospect:PureZprime}.}
  \label{fig:prospect:CombinedZPrimeScalar}
\end{figure}

As a result of the extra production channel of $\varphi \to Z'Z'$ for $Z'$, the improved DUNE sensitivities to the parameter space of $m_{Z'}$ and $g_{BL}$ are shown in Fig.~\ref{fig:prospect:CombinedZPrimeScalar} as the dashed and dot-dashed black lines. As in Fig.~\ref{fig:prospect:PureZprime}, we use ten signal events as the required amount to be statistically significant. The dashed line (reaching lower $g_{BL}$) denotes the extended reach subject to the current constraints on $m_\varphi$ and $\sin^2\vartheta$ from CHARM, KOTO, NA62, and E949 discussed in Section~\ref{subsec:EffectonScalarLims}. The dot-dashed line (extending below the solid black one but not as low as the dashed one) indicates the improvement subject to DUNE's direct $\varphi$ decay search. In effect, if a signal of $Z'$ decay is observed consistent with the parameter space \textit{between} the dot-dashed and dashed lines, then a similar $\varphi \to \ell^+ \ell^-$ or $\varphi \to \pi \pi$ signature should \textit{also} be observable with the same data. Unlike in Fig.~\ref{fig:prospect:PureZprime}, here we do not show the prospects at future FASER and FASER2, because the presence of $\varphi$ could in principle modify these projections and it requires a dedicated FASER simulation\footnote{An analogous combined scalar/vector boson search was explored in the context of dark photons at FASER in Ref.~\cite{Araki:2020wkq}.}, which is beyond the scope of our current work. The other labels are the same as in Fig.~\ref{fig:prospect:PureZprime}. In principle all other limits on $Z'$ such as those from the beam-dump experiments could also be affected by the scalar $\varphi$. However, for simplicity we do not include these corrections and assume na\"{i}vely that they are small.

The two BPs in Eqs.~(\ref{eqn:BP1}) and (\ref{eqn:BP2}) are indicated by the stars in Fig.~\ref{fig:prospect:CombinedZPrimeScalar}. The BP with $m_{Z'} = 140$ MeV is out of all current existing limits and even the $Z'$ prospects at DUNE in the pure gauge boson case in Section~\ref{subsec:PureZPrime}. However, it can be probed at DUNE in presence of the channel $\varphi \to Z'Z'$. As shown in Fig.~\ref{fig:prospect:CombinedZPrimeScalar}, the BP with $m_{Z'} = 30$ MeV is out of the reach of DUNE prospects in the pure $Z'$ case, but can be probed in presence of the scalar $\varphi$. It should be noted that this BP is nearly precluded by the supernova limits on $Z'$~\cite{Croon:2020lrf}. However, as mentioned in Section~\ref{subsec:EffectonScalarLims}, in presence of $\varphi$ the supernova limits on $Z'$ could be very different. Furthermore, the supernova limits on $Z'$ could also change dramatically and thus be possibly avoided if $Z'$ couples to a dark matter particle~\cite{Zhang:2014wra}. In addition, the interplay of $\varphi$ and $Z'$ would also be important for the evolution of $Z'$ in the early universe and the resultant BBN limits in Fig.~\ref{fig:prospect:CombinedZPrimeScalar}.

It is clear in Fig.~\ref{fig:prospect:CombinedZPrimeScalar} that with the extra contribution from the scalar $\varphi$, DUNE can extend its search for $Z'$ to smaller $g_{BL}$ by a factor of 7 when subject to DUNE sensitivity to direct $\varphi$ decays and by a factor of 45 when considering current $\varphi$ constraints. It is striking that the scalar $\varphi$ can improve the $Z'$ prospects at DUNE far beyond the current beam-dump limits for $m_{Z'} \lesssim 100$ MeV, which would be otherwise very challenging for the pure $Z'$ boson case.

\section{Conclusion}
\label{sec:conclusion}

We have explored the possibility of experimentally searching for a class of BSM scenarios where the SM is extended by a minimal $U(1)_{B-L}$ gauge symmetry to explain neutrino masses and possibly provide a portal to DM.  We focused on the light $Z'$ boson and light associated $U(1)_{B-L}$-breaking scalar $\varphi$ in the low gauge coupling regime. We discussed how the presence of $Z'$ and $\varphi$ affects the search range at high-intensity facilities such as the near detector complex at DUNE. Focusing primarily on the sub-GeV mass region for $Z'$ and $\varphi$, the main results of this paper can be summarized as follows:
\begin{itemize}
    \item If the scalar $\varphi$ decouples, the DUNE prospects of pure $Z'$ boson is presented in Fig.~\ref{fig:prospect:PureZprime}. Compared to the FASER sensitivities, the DUNE near detector complex can probe a smaller coupling $g_{BL}$ as a result of the lower center-of-mass proton-beam/target energy at DUNE (compared to the LHC).

    \item As far as the $B-L$ breaking scalar $\varphi$ is concerned, the presence of $Z'$ somewhat weakens the search range of its parameters at DUNE, compared to the case without the gauge interaction, as shown in Fig.~\ref{fig:scalar:varyBr}.

    \item With the extra source of $Z'$ from scalar decay $\varphi \to Z'Z'$, the number of $Z'$ signal events can be greatly increased for $m_{Z'} \lesssim 200$ MeV (cf. Fig.~\ref{fig:CombinedExplanation}), and the DUNE prospects of $g_{BL}$ can be improved by up to a factor of 45, as demonstrated in Fig.~\ref{fig:prospect:CombinedZPrimeScalar}.
\end{itemize}
Our results are largely complementary to the other works on $B-L$ $Z'$ and $\varphi$ searches at DUNE and other facilities such as FASER and SHiP detectors, as well as to the limits from BBN and supernovae.

\section*{Acknowledgement}
We thank Rebecca Leane for clarification of supernova constraints on $B-L$ vector bosons and for providing data points regarding these constraints. The work of P.~S.~B.~D. is supported in part by the U.S. Department of Energy under Grant No.~DE-SC0017987, by the Neutrino Theory Network Program, and by a Fermilab Intensity Frontier Fellowship. The work of B.~D. is supported by the U.S. Department of Energy (DOE) Grant DE-SC0010813. K.~J.~K. is supported by Fermi Research Alliance, LLC, under contract DE-AC02-07CH11359 with the U.S. Department of Energy. The work of R.~N.~M. is supported by the National Science Foundation grant no. PHY-1914631.  Y.~Z. is partially supported by ``the Fundamental Research Funds for the Central Universities''.

\appendix

\section{Details of \texorpdfstring{$\varphi$}{phi} Decay}
\label{app:A}

The transition amplitude $G(s)$ appearing in Eq.~\eqref{eqn:phipipi} (with $s=m_\varphi^2$) consists of three form factors. In the limit of isospin conservation, $G(s)$ can be written in the form of~\cite{Donoghue:1990xh}
\begin{eqnarray}
G(s) = \frac{2}{9} \theta_\pi (s) + \frac79 \left[ \Gamma_\pi (s) + \Delta_\pi (s) \right] \,,
\end{eqnarray}
where in chiral perturbation theory the three form factors are respectively
\begin{eqnarray}
\theta_\pi (s) & = & \left( 1 + \frac{2m_\pi^2}{s} \right) (1+ \psi (s)) + b_\theta s \,, \\
\Gamma_\pi (s) & = &  \frac{m_\pi^2}{s} (1+ \psi (s) + b_\Gamma s )  \,, \\
\Delta_\pi (s) & = &  d_F (1+ \psi (s) + b_\Delta s )  \,,
\end{eqnarray}
with the coefficients $b_\theta = 2.7 \; {\rm GeV}^{-2}$, $b_\Gamma = 2.6 \; {\rm GeV}^{-2}$, $b_\Delta = 3.3 \; {\rm GeV}^{-2}$, $d_F = 0.09$, and the function
\begin{eqnarray}
\psi(s) = \frac{2s - m_\pi^2}{16\pi^2 f_\pi^2}
\left[ \kappa \log \left( \frac{1-\kappa}{1+\kappa} \right) + 2 + i\pi \kappa \right]
+ \frac{s}{96\pi^2 f_\pi^2} \,,
\end{eqnarray}
where $f_\pi = 130$ MeV is the pion decay constant, and $\kappa = \sqrt{1-4m_\pi^2/s}$.

In Eq.~(\ref{eqn:phiAA}) the loop functions are defined as~\cite{Djouadi:2005gi}
\begin{eqnarray}
A_{1/2} (\tau) & \ = \ & 2 \left[ \tau + (\tau-1) f(\tau) \right] \tau^{-2} \,, \\
A_{1} (\tau) & \ = \ & - \left[ 2\tau^2 + 3\tau + 3 (2\tau-1) f(\tau) \right] \tau^{-2} \,,
\end{eqnarray}
with $\tau_X = m_\varphi^2/4m_X^2$, and the function
\begin{eqnarray}
f(\tau) \ = \begin{cases}
\arcsin^2\sqrt{\tau} \,, & \tau \leq 1 \,, \\
-\frac14 \left[ \log \left( \frac{1+ \sqrt{1-1/\tau}}{1-\sqrt{1-1/\tau}} \right) - i\pi \right]^2 \,, & \tau >1 \,.
\end{cases}
\end{eqnarray}

\section{Additional Subdominant Contributions to the Gauge Boson Flux}\label{appendix:ZPrimeFluxes}

For completeness, we list here possible additional production channels of $Z'$ in the $U(1)_{B-L}$ model at DUNE, although they are small with respect to the dominant channels in Section~\ref{subsubsec:ZPrimeProduction}.
\begin{itemize}
    \item Two-body neutral meson decay  $K_L \to \gamma Z$, with the BRs~\cite{Batell:2009di}
\begin{eqnarray}
{\rm BR} (K_L \to \gamma Z')  & \ \simeq \ & 2 {\rm BR} (K_L \to \gamma\gamma) \times \frac{g_{BL}^2}{e^2}
\left( 1 - \frac{m_{Z'}^2}{m_{K^0}^2} \right)^3 \,, 
\end{eqnarray}
  which is highly suppressed by the small SM $ {\rm BR} (K_L \to \gamma\gamma) = 5.47 \times 10^{-4}$~\cite{Zyla:2020zbs}.

    \item Three-body charged pion/Kaon decays $\pi^+ \to \ell^{+} \nu Z'$ and $K^+ \to \ell^+ \nu Z'$ with $\ell = e,\,\mu$. Because the incoming mesons have zero baryon or lepton number, this decay width should be identical to the leptonic $Z'$ case derived in Appendix B of Ref.~\cite{Berryman:2019dme} (see also Ref.~\cite{Krnjaic:2019rsv}). This process can be calculated numerically -- we find that the production of $B-L$ $Z'$ gauge bosons from charged meson decays is suppressed relative to neutral meson decay/bremsstrahlung production, which we considered in the main text.

  \item Loop-level flavor-changing two-body decays $K^+ \to \pi^+ Z'$ and $K_S \to \pi^0 Z'$. Based on the low-energy chiral theory~\cite{Ecker:1987qi, DAmbrosio:1998gur}, the width is~\cite{Pospelov:2008zw, Davoudiasl:2014kua}
\begin{eqnarray}
\label{eqn:KpiZp}
\Gamma (K \to \pi Z') & = & \frac{g_{BL}^2 m_{Z'}^2 W^2 (m_{Z'}^2)}{2^{12} \pi^5 m_K }
\lambda(m_K,\, m_\pi,\, m_{Z'}) \nonumber \\
&& \times \left[ \left( 1 - \frac{m_\pi^2}{m_K^2} \right)^2
+ \frac{m_{Z'}^2}{m_K^2} \left( 2 + \frac{2m_\pi^2}{m_K^2} - \frac{m_{Z'}^2}{m_K^2} \right) \right] \,,
\end{eqnarray}
where $W^2 (q^2)$ incorporates both the polarization and $\pi\pi$ contributions~\cite{DAmbrosio:1998gur}, and
  \begin{eqnarray}
  \lambda (a,b,c) \equiv \left[ 1 - \frac{2(b^2+c^2)}{a^2} + \frac{(b^2-c^2)^2}{a^4} \right]^{1/2} \,.
  \end{eqnarray}
  Note that in the limit of $m_{Z'} \to 0$, the width vanishes, as required by angular momentum conservation. The decay width in Eq.~(\ref{eqn:KpiZp}) applies to both $K^+ \to \pi^+ Z'$ and $K_S \to \pi^0 Z'$, which, however, are highly suppressed by the loop factor:
  \begin{eqnarray}
  {\rm BR} (K^\pm \to \pi^\pm Z') & \ \simeq \ & 8 \times 10^{-4} \times
  \left( \frac{m_{Z'}}{100 \, {\rm MeV}} \right)^2 \lambda(m_{K^\pm},\, m_{\pi^\pm},\, m_{Z'}) \,, \\
  {\rm BR} (K_S \to \pi^0 Z') & \ \simeq \ & 6 \times 10^{-6} \times
  \left( \frac{m_{Z'}}{100 \, {\rm MeV}} \right)^2 \lambda(m_{K^0},\, m_{\pi^0},\, m_{Z'}) \,.
  \end{eqnarray}
  On the contrary, the process $K_L \to \pi \gamma^\ast$ is CP-violating, and as a result the decay $K_L \to \pi^0 Z'$ is highly suppressed~\cite{Ecker:1987qi, DAmbrosio:1998gur}.  There are also the flavor-violating decays $B \to \pi Z'$ and $B \to K Z'$. However, the production rate of $B$ mesons at the DUNE is highly suppressed~\cite{Adams:2013qkq}, and we will not consider these $B$ meson decay channels any more.

  \item Baryon decay $\Delta (1232) \to N Z'$ with $N$ here referring to nucleon, The corresponding BR is~\cite{Batell:2009di} 
\begin{eqnarray}
{\rm Br} (\Delta \to N Z') = {\rm BR} (\Delta \to N \gamma) \times \frac{g_{BL}^2}{e^2}
\left( 1 - \frac{m_{Z'}^2}{(m_\Delta - m_N)^2} \right)^{3/2} \,,
\end{eqnarray}
which is also suppressed by the SM ${\rm BR} ( \Delta \to N \gamma) \simeq 6 \times 10^{-3}$~\cite{Zyla:2020zbs}.

\end{itemize}

\section{Additional \texorpdfstring{$Z'$}{Zp} Boson Constraints}
\label{appendix:Zprimelimit}

The current limits on the light $Z'$ boson of the $U(1)_{B-L}$ model are summarized in Fig.~\ref{fig:prospect:PureZprime}.
Here we discuss a couple more limits from rare meson decays, which are not included in Ref.~\cite{Bauer:2018onh}. The most relevant ones are those from the loop-level flavor-changing $K^+$ decay (see Eq.~(\ref{eqn:KpiZp}) for the width)
\begin{eqnarray}
K^+  \to \pi^+ Z'\,, \quad Z' \to \nu\bar\nu \,.
\end{eqnarray}
The current most stringent limits are from NA62~\cite{Ruggiero:2019} and E949~\cite{Artamonov:2009sz}.  The NA62 experiment has obtained a 95\% C.L. upper limit ${\rm BR} (K^+ \to \pi^+ \nu\bar\nu) < 2.44 \times 10^{-10}$, which can be translated into the limit on $m_{Z'}$ and $g_{BL}$ plane, as shown by the orange line in Fig.~\ref{fig:prospect:PureZprime}. The E949 limit on ${\rm BR} (K^+ \to \pi^+ + X)$ (with $X$ a long-lived particle) can reach up to $5.4 \times 10^{-11}$, which is shown as the purple line in Fig.~\ref{fig:prospect:PureZprime}. The future NA62 data can improve the ${\rm BR} (K^+ \to \pi^+ \nu\bar\nu)$ constraint down to $2.35\times10^{-11}$~\cite{Anelli:2005ju}, and the resulting limit on $g_{BL}$ can be enhanced by a factor of 3, as indicated by the dashed orange line in Fig.~\ref{fig:prospect:PureZprime}.

The gaps in the NA62 and E949 limits are due to the significant background due to the decays $K^+ \to \pi^+ \pi^0$ with $\pi^0 \to \nu\bar\nu$. Note that because the width of $K^+ \to \pi^+ Z'$ is proportional to $Z'$ mass (cf.~Eq.~(\ref{eqn:KpiZp})), both the NA62 and E949 limits get weaker when $Z'$ is lighter. 
There are in principle also the limits for a long-lived $Z'$, with $Z'$ decaying outside the detectors. However, to have a long-lived $Z'$ the coupling $g_{BL}$ needs to be very small, which will in turn highly suppress the production of $Z'$. It turns out that no parameter space of $m_{Z'}$ and $g_{BL}$ is excluded due to the long-lived $Z'$ at NA62 and E949. The KOTO experiment has put an upper bound of ${\rm BR} (K_L \to \pi \nu\bar\nu) < 3.0 \times 10^{-9}$~\cite{Ahn:2018mvc, Ahn:2020opg}. However, this limit can not be used to set constraints on the decay $K_S \to \pi^0 Z'$.


There are more limits from the meson and baryon decays which can be used to constrain $m_{Z'}$ and $g_{BL}$, which are however much weaker, thus are not shown in Fig.~\ref{fig:prospect:PureZprime}:
\begin{itemize}
  \item The two-body meson decay $K_L \to \gamma Z'$ with the subsequent decays $Z' \to e^+ e^-,\, \mu^+ \mu^-$ contributes respectively to the decays $K_L \to e^+ e^- \gamma$~\cite{Ohl:1990qw, Barr:1990zh, Fanti:1999rz} and $K_L \to \mu^+ \mu^- \gamma$~\cite{AlaviHarati:2001wd}. However, as stated in Appendix~\ref{appendix:ZPrimeFluxes}, the meson decay $K_L \to \gamma Z'$ is highly suppressed by the corresponding SM ${\rm BR} (K_L \to \gamma\gamma)$.

  \item The baryon decay $\Delta \to p Z'$ with $Z' \to e^+ e^-$ contributes to $\Delta \to p e^+ e^-$, which is highly suppressed by the small SM ${\rm BR} (\Delta \to N\gamma)$, as mentioned in Appendix~\ref{appendix:ZPrimeFluxes}.

\end{itemize}

\bibliography{ref}

\providecommand{\href}[2]{#2}\begingroup\raggedright\begin{thebibliography}{100}

\bibitem{Davidson:1978pm}
A.~Davidson, {\it {$B-L$ as the fourth color within an $SU(2)_L \times U(1)_R
  \times U(1)$ model}},  {\em Phys. Rev.} {\bf D20} (1979) 776.

\bibitem{Marshak:1979fm}
R.~E. Marshak and R.~N. Mohapatra, {\it {Quark - Lepton Symmetry and $B-L$ as
  the $U(1)$ Generator of the Electroweak Symmetry Group}},  {\em Phys. Lett.}
  {\bf 91B} (1980) 222--224.

\bibitem{Minkowski:1977sc}
P.~Minkowski, {\it {$\mu \to e\gamma$ at a Rate of One Out of $10^{9}$ Muon
  Decays?}},  {\em Phys. Lett. B} {\bf 67} (1977) 421--428.

\bibitem{Mohapatra:1979ia}
R.~N. Mohapatra and G.~Senjanovic, {\it {Neutrino Mass and Spontaneous Parity
  Nonconservation}},  {\em Phys. Rev. Lett.} {\bf 44} (1980) 912.

\bibitem{Yanagida:1979as}
T.~Yanagida, {\it {Horizontal gauge symmetry and masses of neutrinos}},  {\em
  Conf. Proc. C} {\bf 7902131} (1979) 95--99.

\bibitem{GellMann:1980vs}
M.~Gell-Mann, P.~Ramond, and R.~Slansky, {\it {Complex Spinors and Unified
  Theories}},  {\em Conf. Proc. C} {\bf 790927} (1979) 315--321,
  [\href{http://www.arxiv.org/abs/1306.4669}{{\tt 1306.4669}}].

\bibitem{Glashow:1979nm}
S.~Glashow, {\it {The Future of Elementary Particle Physics}},  {\em NATO Sci.
  Ser. B} {\bf 61} (1980) 687.

\bibitem{Mohapatra:2019ysk}
R.~N. Mohapatra and N.~Okada, {\it {Dark Matter Constraints on Low Mass and
  Weakly Coupled B-L Gauge Boson}},  {\em Phys. Rev. D} {\bf 102} (2020), no.~3
  035028, [\href{http://www.arxiv.org/abs/1908.11325}{{\tt 1908.11325}}].

\bibitem{Brehmer:2015cia}
J.~Brehmer, J.~Hewett, J.~Kopp, T.~Rizzo, and J.~Tattersall, {\it {Symmetry
  Restored in Dibosons at the LHC?}},  {\em JHEP} {\bf 10} (2015) 182,
  [\href{http://www.arxiv.org/abs/1507.00013}{{\tt 1507.00013}}].

\bibitem{Dev:2016dja}
P.~S.~B. Dev, R.~N. Mohapatra, and Y.~Zhang, {\it {Probing the Higgs Sector of
  the Minimal Left-Right Symmetric Model at Future Hadron Colliders}},  {\em
  JHEP} {\bf 05} (2016) 174, [\href{http://www.arxiv.org/abs/1602.05947}{{\tt
  1602.05947}}].

\bibitem{Chauhan:2018uuy}
G.~Chauhan, P.~S.~B. Dev, R.~N. Mohapatra, and Y.~Zhang, {\it {Perturbativity
  constraints on $U(1)_{B-L}$ and left-right models and implications for heavy
  gauge boson searches}},  {\em JHEP} {\bf 01} (2019) 208,
  [\href{http://www.arxiv.org/abs/1811.08789}{{\tt 1811.08789}}].

\bibitem{Okada:2012sg}
N.~Okada and Y.~Orikasa, {\it {Dark matter in the classically conformal B-L
  model}},  {\em Phys. Rev. D} {\bf 85} (2012) 115006,
  [\href{http://www.arxiv.org/abs/1202.1405}{{\tt 1202.1405}}].

\bibitem{Kaneta:2016vkq}
K.~Kaneta, Z.~Kang, and H.-S. Lee, {\it {Right-handed neutrino dark matter
  under the $B - L$ gauge interaction}},  {\em JHEP} {\bf 02} (2017) 031,
  [\href{http://www.arxiv.org/abs/1606.09317}{{\tt 1606.09317}}].

\bibitem{Klasen:2016qux}
M.~Klasen, F.~Lyonnet, and F.~S. Queiroz, {\it {NLO+NLL collider bounds, Dirac
  fermion and scalar dark matter in the $B-L$ model}},  {\em Eur. Phys. J.}
  {\bf C77} (2017), no.~5 348, [\href{http://www.arxiv.org/abs/1607.06468}{{\tt
  1607.06468}}].

\bibitem{Heeba:2019jho}
S.~Heeba and F.~Kahlhoefer, {\it {Probing the freeze-in mechanism in dark
  matter models with $U(1)'$ gauge extensions}},  {\em Phys. Rev. D} {\bf 101}
  (2020), no.~3 035043, [\href{http://www.arxiv.org/abs/1908.09834}{{\tt
  1908.09834}}].

\bibitem{Mohapatra:2020bze}
R.~N. Mohapatra and N.~Okada, {\it {Freeze-in Dark Matter from a Minimal B-L
  Model and Possible Grand Unification}},  {\em Phys. Rev. D} {\bf 101} (2020),
  no.~11 115022, [\href{http://www.arxiv.org/abs/2005.00365}{{\tt
  2005.00365}}].

\bibitem{Borah:2020wyc}
D.~Borah, S.~Jyoti~Das, and A.~K. Saha, {\it {Cosmic inflation in minimal
  $U(1)_{B-L}$ model: implications for (non) thermal dark matter and
  leptogenesis}},  {\em Eur. Phys. J. C} {\bf 81} (2021), no.~2 169,
  [\href{http://www.arxiv.org/abs/2005.11328}{{\tt 2005.11328}}].

\bibitem{Wetterich:1981bx}
C.~Wetterich, {\it {Neutrino Masses and the Scale of B-L Violation}},  {\em
  Nucl. Phys.} {\bf B187} (1981) 343--375.

\bibitem{Buchmuller:1991ce}
W.~Buchmuller, C.~Greub, and P.~Minkowski, {\it {Neutrino masses, neutral
  vector bosons and the scale of $B-L$ breaking}},  {\em Phys. Lett.} {\bf
  B267} (1991) 395--399.

\bibitem{Emam:2007dy}
W.~Emam and S.~Khalil, {\it {Higgs and $Z^\prime$ phenomenology in B-L
  extension of the standard model at LHC}},  {\em Eur. Phys. J.} {\bf C52}
  (2007) 625--633, [\href{http://www.arxiv.org/abs/0704.1395}{{\tt
  0704.1395}}].

\bibitem{Basso:2008iv}
L.~Basso, A.~Belyaev, S.~Moretti, and C.~H. Shepherd-Themistocleous, {\it
  {Phenomenology of the minimal $B-L$ extension of the Standard model:
  $Z^\prime$ and neutrinos}},  {\em Phys. Rev.} {\bf D80} (2009) 055030,
  [\href{http://www.arxiv.org/abs/0812.4313}{{\tt 0812.4313}}].

\bibitem{Perez:2009mu}
P.~Fileviez~Perez, T.~Han, and T.~Li, {\it {Testability of Type I Seesaw at the
  CERN LHC: Revealing the Existence of the B-L Symmetry}},  {\em Phys. Rev.}
  {\bf D80} (2009) 073015, [\href{http://www.arxiv.org/abs/0907.4186}{{\tt
  0907.4186}}].

\bibitem{Basso:2010jm}
L.~Basso, S.~Moretti, and G.~M. Pruna, {\it {A Renormalisation Group Equation
  Study of the Scalar Sector of the Minimal B-L Extension of the Standard
  Model}},  {\em Phys. Rev.} {\bf D82} (2010) 055018,
  [\href{http://www.arxiv.org/abs/1004.3039}{{\tt 1004.3039}}].

\bibitem{Heeck:2014zfa}
J.~Heeck, {\it {Unbroken $B-L$ symmetry}},  {\em Phys. Lett. B} {\bf 739}
  (2014) 256--262, [\href{http://www.arxiv.org/abs/1408.6845}{{\tt
  1408.6845}}].

\bibitem{Khalil:2006yi}
S.~Khalil, {\it {Low scale $B-L$ extension of the Standard Model at the LHC}},
  {\em J. Phys. G} {\bf 35} (2008) 055001,
  [\href{http://www.arxiv.org/abs/hep-ph/0611205}{{\tt hep-ph/0611205}}].

\bibitem{Huitu:2008gf}
K.~Huitu, S.~Khalil, H.~Okada, and S.~K. Rai, {\it {Signatures for right-handed
  neutrinos at the Large Hadron Collider}},  {\em Phys. Rev. Lett.} {\bf 101}
  (2008) 181802, [\href{http://www.arxiv.org/abs/0803.2799}{{\tt 0803.2799}}].

\bibitem{Accomando:2016sge}
E.~Accomando, C.~Coriano, L.~Delle~Rose, J.~Fiaschi, C.~Marzo, and S.~Moretti,
  {\it {$Z^\prime$, Higgses and heavy neutrinos in $U(1)^\prime$ models: from
  the LHC to the GUT scale}},  {\em JHEP} {\bf 07} (2016) 086,
  [\href{http://www.arxiv.org/abs/1605.02910}{{\tt 1605.02910}}].

\bibitem{Accomando:2017qcs}
E.~Accomando, L.~Delle~Rose, S.~Moretti, E.~Olaiya, and C.~H.
  Shepherd-Themistocleous, {\it {Extra Higgs boson and $Z^\prime$ as portals to
  signatures of heavy neutrinos at the LHC}},  {\em JHEP} {\bf 02} (2018) 109,
  [\href{http://www.arxiv.org/abs/1708.03650}{{\tt 1708.03650}}].

\bibitem{Dev:2017xry}
P.~S.~B. Dev, R.~N. Mohapatra, and Y.~Zhang, {\it {Leptogenesis constraints on
  $B - L$ breaking Higgs boson in TeV scale seesaw models}},  {\em JHEP} {\bf
  03} (2018) 122, [\href{http://www.arxiv.org/abs/1711.07634}{{\tt
  1711.07634}}].

\bibitem{Alioli:2017nzr}
S.~Alioli, M.~Farina, D.~Pappadopulo, and J.~T. Ruderman, {\it {Catching a New
  Force by the Tail}},  {\em Phys. Rev. Lett.} {\bf 120} (2018), no.~10 101801,
  [\href{http://www.arxiv.org/abs/1712.02347}{{\tt 1712.02347}}].

\bibitem{Berryman:2019dme}
J.~M. Berryman, A.~de~Gouvea, P.~J. Fox, B.~J. Kayser, K.~J. Kelly, and J.~L.
  Raaf, {\it {Searches for Decays of New Particles in the DUNE Multi-Purpose
  Near Detector}},  {\em JHEP} {\bf 02} (2020) 174,
  [\href{http://www.arxiv.org/abs/1912.07622}{{\tt 1912.07622}}].

\bibitem{Batell:2009di}
B.~Batell, M.~Pospelov, and A.~Ritz, {\it {Exploring Portals to a Hidden Sector
  Through Fixed Targets}},  {\em Phys. Rev.} {\bf D80} (2009) 095024,
  [\href{http://www.arxiv.org/abs/0906.5614}{{\tt 0906.5614}}].

\bibitem{Blumlein:2013cua}
J.~Bl{\"u}mlein and J.~Brunner, {\it {New Exclusion Limits on Dark Gauge Forces
  from Proton Bremsstrahlung in Beam-Dump Data}},  {\em Phys. Lett.} {\bf B731}
  (2014) 320--326, [\href{http://www.arxiv.org/abs/1311.3870}{{\tt
  1311.3870}}].

\bibitem{deNiverville:2016rqh}
P.~deNiverville, C.-Y. Chen, M.~Pospelov, and A.~Ritz, {\it {Light dark matter
  in neutrino beams: production modelling and scattering signatures at
  MiniBooNE, T2K and SHiP}},  {\em Phys. Rev. D} {\bf 95} (2017), no.~3 035006,
  [\href{http://www.arxiv.org/abs/1609.01770}{{\tt 1609.01770}}].

\bibitem{Buschmann:2015awa}
M.~Buschmann, J.~Kopp, J.~Liu, and P.~A.~N. Machado, {\it {Lepton Jets from
  Radiating Dark Matter}},  {\em JHEP} {\bf 07} (2015) 045,
  [\href{http://www.arxiv.org/abs/1505.07459}{{\tt 1505.07459}}].

\bibitem{Ilten:2018crw}
P.~Ilten, Y.~Soreq, M.~Williams, and W.~Xue, {\it {Serendipity in dark photon
  searches}},  {\em JHEP} {\bf 06} (2018) 004,
  [\href{http://www.arxiv.org/abs/1801.04847}{{\tt 1801.04847}}].

\bibitem{Bauer:2018onh}
M.~Bauer, P.~Foldenauer, and J.~Jaeckel, {\it {Hunting All the Hidden
  Photons}},  {\em JHEP} {\bf 07} (2018) 094,
  [\href{http://www.arxiv.org/abs/1803.05466}{{\tt 1803.05466}}].
  [JHEP18,094(2020)].

\bibitem{Ballett:2019xoj}
P.~Ballett, M.~Hostert, S.~Pascoli, Y.~F. Perez-Gonzalez, Z.~Tabrizi, and
  R.~Zukanovich~Funchal, {\it {$Z^\prime$s in neutrino scattering at DUNE}},
  {\em Phys. Rev. D} {\bf 100} (2019), no.~5 055012,
  [\href{http://www.arxiv.org/abs/1902.08579}{{\tt 1902.08579}}].

\bibitem{Altmannshofer:2019zhy}
W.~Altmannshofer, S.~Gori, J.~Mart\'\i{}n-Albo, A.~Sousa, and M.~Wallbank, {\it
  {Neutrino Tridents at DUNE}},  {\em Phys. Rev. D} {\bf 100} (2019), no.~11
  115029, [\href{http://www.arxiv.org/abs/1902.06765}{{\tt 1902.06765}}].

\bibitem{Batell:2009jf}
B.~Batell, M.~Pospelov, and A.~Ritz, {\it {Multi-lepton Signatures of a Hidden
  Sector in Rare B Decays}},  {\em Phys. Rev.} {\bf D83} (2011) 054005,
  [\href{http://www.arxiv.org/abs/0911.4938}{{\tt 0911.4938}}].

\bibitem{Acciarri:2015uup}
{\bf DUNE} {\bf Collaboration}, R.~Acciarri {\em et~al.}, {\it {Long-Baseline
  Neutrino Facility (LBNF) and Deep Underground Neutrino Experiment (DUNE)}:
  {Conceptual Design Report, Volume 2: The Physics Program for DUNE at LBNF}},
  \href{http://www.arxiv.org/abs/1512.06148}{{\tt 1512.06148}}.

\bibitem{Acciarri:2016crz}
{\bf DUNE} {\bf Collaboration}, R.~Acciarri {\em et~al.}, {\it {Long-Baseline
  Neutrino Facility (LBNF) and Deep Underground Neutrino Experiment (DUNE)}:
  {Conceptual Design Report, Volume 1: The LBNF and DUNE Projects}},
  \href{http://www.arxiv.org/abs/1601.05471}{{\tt 1601.05471}}.

\bibitem{AbedAbud:2021hpb}
{\bf DUNE} {\bf Collaboration}, A.~Abed~Abud {\em et~al.}, {\it {Deep
  Underground Neutrino Experiment (DUNE) Near Detector Conceptual Design
  Report}},  \href{http://www.arxiv.org/abs/2103.13910}{{\tt 2103.13910}}.

\bibitem{Ballett:2019bgd}
P.~Ballett, T.~Boschi, and S.~Pascoli, {\it {Heavy Neutral Leptons from
  low-scale seesaws at the DUNE Near Detector}},  {\em JHEP} {\bf 03} (2020)
  111, [\href{http://www.arxiv.org/abs/1905.00284}{{\tt 1905.00284}}].

\bibitem{Coloma:2020lgy}
P.~Coloma, E.~Fern\'andez-Mart\'\i{}nez, M.~Gonz\'alez-L\'opez,
  J.~Hern\'andez-Garc\'\i{}a, and Z.~Pavlovic, {\it {GeV-scale neutrinos:
  interactions with mesons and DUNE sensitivity}},  {\em Eur. Phys. J. C} {\bf
  81} (2021), no.~1 78, [\href{http://www.arxiv.org/abs/2007.03701}{{\tt
  2007.03701}}].

\bibitem{Kelly:2020dda}
K.~J. Kelly, S.~Kumar, and Z.~Liu, {\it {Heavy Axion Opportunities at the DUNE
  Near Detector}},  \href{http://www.arxiv.org/abs/2011.05995}{{\tt
  2011.05995}}.

\bibitem{Brdar:2020dpr}
V.~Brdar, B.~Dutta, W.~Jang, D.~Kim, I.~M. Shoemaker, Z.~Tabrizi, A.~Thompson,
  and J.~Yu, {\it {Axion-like Particles at Future Neutrino Experiments: Closing
  the "Cosmological Triangle"}},
  \href{http://www.arxiv.org/abs/2011.07054}{{\tt 2011.07054}}.

\bibitem{Dev:2021ofc}
P.~S.~B. Dev, D.~Kim, K.~Sinha, and Y.~Zhang, {\it {PASSAT at Future Neutrino
  Experiments: Hybrid Beam-Dump-Helioscope Facilities to Probe Light Axion-Like
  Particles}},  \href{http://www.arxiv.org/abs/2101.08781}{{\tt 2101.08781}}.

\bibitem{Breitbach:2021gvv}
M.~Breitbach, L.~Buonocore, C.~Frugiuele, J.~Kopp, and L.~Mittnacht, {\it
  {Searching for Physics Beyond the Standard Model in an Off-Axis DUNE Near
  Detector}},  \href{http://www.arxiv.org/abs/2102.03383}{{\tt 2102.03383}}.

\bibitem{Bakhti:2018avv}
P.~Bakhti, Y.~Farzan, and M.~Rajaee, {\it {Secret interactions of neutrinos
  with light gauge boson at the DUNE near detector}},  {\em Phys. Rev.} {\bf
  D99} (2019), no.~5 055019, [\href{http://www.arxiv.org/abs/1810.04441}{{\tt
  1810.04441}}].

\bibitem{Deniz:2009mu}
{\bf TEXONO} {\bf Collaboration}, M.~Deniz {\em et~al.}, {\it {Measurement of
  Nu(e)-bar -Electron Scattering Cross-Section with a CsI(Tl) Scintillating
  Crystal Array at the Kuo-Sheng Nuclear Power Reactor}},  {\em Phys. Rev. D}
  {\bf 81} (2010) 072001, [\href{http://www.arxiv.org/abs/0911.1597}{{\tt
  0911.1597}}].

\bibitem{Bilmis:2015lja}
S.~Bilmis, I.~Turan, T.~M. Aliev, M.~Deniz, L.~Singh, and H.~T. Wong, {\it
  {Constraints on Dark Photon from Neutrino-Electron Scattering Experiments}},
  {\em Phys. Rev. D} {\bf 92} (2015), no.~3 033009,
  [\href{http://www.arxiv.org/abs/1502.07763}{{\tt 1502.07763}}].

\bibitem{Bergsma:1985qz}
{\bf CHARM} {\bf Collaboration}, F.~Bergsma {\em et~al.}, {\it {Search for
  Axion Like Particle Production in 400-GeV Proton - Copper Interactions}},
  {\em Phys. Lett. B} {\bf 157} (1985) 458--462.

\bibitem{Lees:2014xha}
{\bf BaBar} {\bf Collaboration}, J.~P. Lees {\em et~al.}, {\it {Search for a
  Dark Photon in $e^+e^-$ Collisions at BaBar}},  {\em Phys. Rev. Lett.} {\bf
  113} (2014), no.~20 201801, [\href{http://www.arxiv.org/abs/1406.2980}{{\tt
  1406.2980}}].

\bibitem{Artamonov:2009sz}
{\bf BNL-E949} {\bf Collaboration}, A.~V. Artamonov {\em et~al.}, {\it {Study
  of the decay $K^+\to\pi^+\nu \bar\nu$ in the momentum region $140 < P_\pi <
  199$ MeV/c}},  {\em Phys. Rev.} {\bf D79} (2009) 092004,
  [\href{http://www.arxiv.org/abs/0903.0030}{{\tt 0903.0030}}].

\bibitem{Ruggiero:2019}
G.~Ruggiero, {\it Latest measurement of $k^+ \to \pi^+ \nu\bar\nu$ with the
  na62 experiment at cern}, . Talk given at KAON2019, Perugia, Italy.

\bibitem{Knapen:2017xzo}
S.~Knapen, T.~Lin, and K.~M. Zurek, {\it {Light Dark Matter: Models and
  Constraints}},  {\em Phys. Rev.} {\bf D96} (2017), no.~11 115021,
  [\href{http://www.arxiv.org/abs/1709.07882}{{\tt 1709.07882}}].

\bibitem{Croon:2020lrf}
D.~Croon, G.~Elor, R.~K. Leane, and S.~D. McDermott, {\it {Supernova Muons: New
  Constraints on $Z$' Bosons, Axions and ALPs}},  {\em JHEP} {\bf 01} (2021)
  107, [\href{http://www.arxiv.org/abs/2006.13942}{{\tt 2006.13942}}].

\bibitem{Bennett:2006fi}
{\bf Muon g-2} {\bf Collaboration}, G.~W. Bennett {\em et~al.}, {\it {Final
  Report of the Muon E821 Anomalous Magnetic Moment Measurement at BNL}},  {\em
  Phys. Rev. D} {\bf 73} (2006) 072003,
  [\href{http://www.arxiv.org/abs/hep-ex/0602035}{{\tt hep-ex/0602035}}].

\bibitem{Abi:2021gix}
{\bf Muon g-2} {\bf Collaboration}, B.~Abi {\em et~al.}, {\it {Measurement of
  the Positive Muon Anomalous Magnetic Moment to 0.46 ppm}},  {\em Phys. Rev.
  Lett.} {\bf 126} (2021), no.~14 141801,
  [\href{http://www.arxiv.org/abs/2104.03281}{{\tt 2104.03281}}].

\bibitem{Anelli:2005ju}
G.~Anelli {\em et~al.}, {\it {Proposal to measure the rare decay $K^+ \to \pi^+
  \nu \bar{\nu}$ at the CERN SPS}}, . CERN-SPSC-2005-013, CERN-SPSC-P-326.

\bibitem{Ariga:2018uku}
{\bf FASER} {\bf Collaboration}, A.~Ariga {\em et~al.}, {\it
  {FASER\textquoteright{}s physics reach for long-lived particles}},  {\em
  Phys. Rev. D} {\bf 99} (2019), no.~9 095011,
  [\href{http://www.arxiv.org/abs/1811.12522}{{\tt 1811.12522}}].

\bibitem{Riordan:1987aw}
E.~M. Riordan {\em et~al.}, {\it {A Search for Short Lived Axions in an
  Electron Beam Dump Experiment}},  {\em Phys. Rev. Lett.} {\bf 59} (1987) 755.

\bibitem{Bjorken:1988as}
J.~D. Bjorken, S.~Ecklund, W.~R. Nelson, A.~Abashian, C.~Church, B.~Lu, L.~W.
  Mo, T.~A. Nunamaker, and P.~Rassmann, {\it {Search for Neutral Metastable
  Penetrating Particles Produced in the SLAC Beam Dump}},  {\em Phys. Rev.}
  {\bf D38} (1988) 3375.

\bibitem{Bjorken:2009mm}
J.~D. Bjorken, R.~Essig, P.~Schuster, and N.~Toro, {\it {New Fixed-Target
  Experiments to Search for Dark Gauge Forces}},  {\em Phys. Rev.} {\bf D80}
  (2009) 075018, [\href{http://www.arxiv.org/abs/0906.0580}{{\tt 0906.0580}}].

\bibitem{Andreas:2012mt}
S.~Andreas, C.~Niebuhr, and A.~Ringwald, {\it {New Limits on Hidden Photons
  from Past Electron Beam Dumps}},  {\em Phys. Rev.} {\bf D86} (2012) 095019,
  [\href{http://www.arxiv.org/abs/1209.6083}{{\tt 1209.6083}}].

\bibitem{Davier:1989wz}
M.~Davier and H.~Nguyen~Ngoc, {\it {An Unambiguous Search for a Light Higgs
  Boson}},  {\em Phys. Lett.} {\bf B229} (1989) 150--155.

\bibitem{Blumlein:2011mv}
J.~Blumlein and J.~Brunner, {\it {New Exclusion Limits for Dark Gauge Forces
  from Beam-Dump Data}},  {\em Phys. Lett.} {\bf B701} (2011) 155--159,
  [\href{http://www.arxiv.org/abs/1104.2747}{{\tt 1104.2747}}].

\bibitem{Tsai:2019mtm}
Y.-D. Tsai, P.~deNiverville, and M.~X. Liu, {\it {The High-Energy Frontier of
  the Intensity Frontier: Closing the Dark Photon, Inelastic Dark Matter, and
  Muon $g-2$ Windows}},  \href{http://www.arxiv.org/abs/1908.07525}{{\tt
  1908.07525}}.

\bibitem{Athanassopoulos:1997er}
{\bf LSND} {\bf Collaboration}, C.~Athanassopoulos {\em et~al.}, {\it {Evidence
  for muon-neutrino $\to$ electron-neutrino oscillations from pion decay in
  flight neutrinos}},  {\em Phys. Rev. C} {\bf 58} (1998) 2489--2511,
  [\href{http://www.arxiv.org/abs/nucl-ex/9706006}{{\tt nucl-ex/9706006}}].

\bibitem{Batell:2019nwo}
B.~Batell, J.~Berger, and A.~Ismail, {\it {Probing the Higgs Portal at the
  Fermilab Short-Baseline Neutrino Experiments}},  {\em Phys. Rev.} {\bf D100}
  (2019), no.~11 115039, [\href{http://www.arxiv.org/abs/1909.11670}{{\tt
  1909.11670}}].

\bibitem{Abe:2010gxa}
{\bf Belle-II} {\bf Collaboration}, T.~Abe {\em et~al.}, {\it {Belle II
  Technical Design Report}},  \href{http://www.arxiv.org/abs/1011.0352}{{\tt
  1011.0352}}.

\bibitem{Kou:2018nap}
{\bf Belle-II} {\bf Collaboration}, W.~Altmannshofer {\em et~al.}, {\it {The
  Belle II Physics Book}},  {\em PTEP} {\bf 2019} (2019), no.~12 123C01,
  [\href{http://www.arxiv.org/abs/1808.10567}{{\tt 1808.10567}}]. [Erratum:
  PTEP 2020, 029201 (2020)].

\bibitem{Aubert:2009cp}
{\bf BaBar} {\bf Collaboration}, B.~Aubert {\em et~al.}, {\it {Search for
  Dimuon Decays of a Light Scalar Boson in Radiative Transitions Upsilon $\to$
  gamma A0}},  {\em Phys. Rev. Lett.} {\bf 103} (2009) 081803,
  [\href{http://www.arxiv.org/abs/0905.4539}{{\tt 0905.4539}}].

\bibitem{Lindner:2016bgg}
M.~Lindner, M.~Platscher, and F.~S. Queiroz, {\it {A Call for New Physics : The
  Muon Anomalous Magnetic Moment and Lepton Flavor Violation}},  {\em Phys.
  Rept.} {\bf 731} (2018) 1--82,
  [\href{http://www.arxiv.org/abs/1610.06587}{{\tt 1610.06587}}].

\bibitem{Borsanyi:2020mff}
S.~Borsanyi {\em et~al.}, {\it {Leading hadronic contribution to the muon 2
  magnetic moment from lattice QCD}},
  \href{http://www.arxiv.org/abs/2002.12347}{{\tt 2002.12347}}.

\bibitem{Aoyama:2020ynm}
T.~Aoyama {\em et~al.}, {\it {The anomalous magnetic moment of the muon in the
  Standard Model}},  {\em Phys. Rept.} {\bf 887} (2020) 1--166,
  [\href{http://www.arxiv.org/abs/2006.04822}{{\tt 2006.04822}}].

\bibitem{Ma:2001md}
E.~Ma, D.~P. Roy, and S.~Roy, {\it {Gauged L(mu) - L(tau) with large muon
  anomalous magnetic moment and the bimaximal mixing of neutrinos}},  {\em
  Phys. Lett. B} {\bf 525} (2002) 101--106,
  [\href{http://www.arxiv.org/abs/hep-ph/0110146}{{\tt hep-ph/0110146}}].

\bibitem{Altmannshofer:2016oaq}
W.~Altmannshofer, M.~Carena, and A.~Crivellin, {\it {$L_\mu - L_\tau$ theory of
  Higgs flavor violation and $(g-2)_\mu$}},  {\em Phys. Rev. D} {\bf 94}
  (2016), no.~9 095026, [\href{http://www.arxiv.org/abs/1604.08221}{{\tt
  1604.08221}}].

\bibitem{Altmannshofer:2016brv}
W.~Altmannshofer, C.-Y. Chen, P.~S.~B. Dev, and A.~Soni, {\it {Lepton flavor
  violating $Z'$ explanation of the muon anomalous magnetic moment}},  {\em
  Phys. Lett. B} {\bf 762} (2016) 389--398,
  [\href{http://www.arxiv.org/abs/1607.06832}{{\tt 1607.06832}}].

\bibitem{CarcamoHernandez:2019ydc}
A.~E. C\'arcamo~Hern\'andez, S.~F. King, H.~Lee, and S.~J. Rowley, {\it {Is it
  possible to explain the muon and electron $g-2$ in a $Z'$ model?}},  {\em
  Phys. Rev. D} {\bf 101} (2020), no.~11 115016,
  [\href{http://www.arxiv.org/abs/1910.10734}{{\tt 1910.10734}}].

\bibitem{Dev:2020drf}
P.~S.~B. Dev, W.~Rodejohann, X.-J. Xu, and Y.~Zhang, {\it {MUonE sensitivity to
  new physics explanations of the muon anomalous magnetic moment}},  {\em JHEP}
  {\bf 05} (2020) 053, [\href{http://www.arxiv.org/abs/2002.04822}{{\tt
  2002.04822}}].

\bibitem{Abdallah:2020biq}
W.~Abdallah, R.~Gandhi, and S.~Roy, {\it {Understanding the MiniBooNE and the
  muon and electron $g - 2$ anomalies with a light $Z'$ and a second Higgs
  doublet}},  {\em JHEP} {\bf 12} (2020) 188,
  [\href{http://www.arxiv.org/abs/2006.01948}{{\tt 2006.01948}}].

\bibitem{Bodas:2021fsy}
A.~Bodas, R.~Coy, and S.~J.~D. King, {\it {Solving the electron and muon $g-2$
  anomalies in $Z'$ models}},  \href{http://www.arxiv.org/abs/2102.07781}{{\tt
  2102.07781}}.

\bibitem{Amaral:2021rzw}
D.~W.~P. Amaral, D.~G. Cerde\~no, A.~Cheek, and P.~Foldenauer, {\it
  {Distinguishing $U(1)_{L_\mu-L_{\tau}}$ from $U(1)_{L_\mu}$ as a solution for
  $(g-2)_\mu$ with neutrinos}},
  \href{http://www.arxiv.org/abs/2104.03297}{{\tt 2104.03297}}.

\bibitem{Athron:2021iuf}
P.~Athron, C.~Bal\'azs, D.~H. Jacob, W.~Kotlarski, D.~St\"ockinger, and
  H.~St\"ockinger-Kim, {\it {New physics explanations of $a_\mu$ in light of
  the FNAL muon $g-2$ measurement}},
  \href{http://www.arxiv.org/abs/2104.03691}{{\tt 2104.03691}}.

\bibitem{Aghanim:2018eyx}
{\bf Planck} {\bf Collaboration}, N.~Aghanim {\em et~al.}, {\it {Planck 2018
  results. VI. Cosmological parameters}},  {\em Astron. Astrophys.} {\bf 641}
  (2020) A6, [\href{http://www.arxiv.org/abs/1807.06209}{{\tt 1807.06209}}].

\bibitem{Escudero:2018mvt}
M.~Escudero, {\it {Neutrino decoupling beyond the Standard Model: CMB
  constraints on the Dark Matter mass with a fast and precise $N_{\rm eff}$
  evaluation}},  {\em JCAP} {\bf 02} (2019) 007,
  [\href{http://www.arxiv.org/abs/1812.05605}{{\tt 1812.05605}}].

\bibitem{Escudero:2020dfa}
M.~Escudero~Abenza, {\it {Precision early universe thermodynamics made simple:
  $N_{\rm eff}$ and neutrino decoupling in the Standard Model and beyond}},
  {\em JCAP} {\bf 05} (2020) 048,
  [\href{http://www.arxiv.org/abs/2001.04466}{{\tt 2001.04466}}].

\bibitem{Kamada:2018zxi}
A.~Kamada, K.~Kaneta, K.~Yanagi, and H.-B. Yu, {\it {Self-interacting dark
  matter and muon $g-2$ in a gauged U$(1)_{L_{\mu} - L_{\tau}}$ model}},  {\em
  JHEP} {\bf 06} (2018) 117, [\href{http://www.arxiv.org/abs/1805.00651}{{\tt
  1805.00651}}].

\bibitem{Escudero:2019gzq}
M.~Escudero, D.~Hooper, G.~Krnjaic, and M.~Pierre, {\it {Cosmology with A Very
  Light L$_{\mu}$ \ensuremath{-} L$_{\tau}$ Gauge Boson}},  {\em JHEP} {\bf 03}
  (2019) 071, [\href{http://www.arxiv.org/abs/1901.02010}{{\tt 1901.02010}}].

\bibitem{Dutta:2020jsy}
B.~Dutta, S.~Ghosh, and J.~Kumar, {\it {Contributions to $\Delta N_{eff}$ from
  the dark photon of $U(1)_{T3R}$}},  {\em Phys. Rev. D} {\bf 102} (2020),
  no.~1 015013, [\href{http://www.arxiv.org/abs/2002.01137}{{\tt 2002.01137}}].

\bibitem{Chang:2016ntp}
J.~H. Chang, R.~Essig, and S.~D. McDermott, {\it {Revisiting Supernova 1987A
  Constraints on Dark Photons}},  {\em JHEP} {\bf 01} (2017) 107,
  [\href{http://www.arxiv.org/abs/1611.03864}{{\tt 1611.03864}}].

\bibitem{Rrapaj:2015wgs}
E.~Rrapaj and S.~Reddy, {\it {Nucleon-nucleon bremsstrahlung of dark gauge
  bosons and revised supernova constraints}},  {\em Phys. Rev. C} {\bf 94}
  (2016), no.~4 045805, [\href{http://www.arxiv.org/abs/1511.09136}{{\tt
  1511.09136}}].

\bibitem{Nelson:2008tn}
A.~E. Nelson and J.~Walsh, {\it {Chameleon vector bosons}},  {\em Phys. Rev. D}
  {\bf 77} (2008) 095006, [\href{http://www.arxiv.org/abs/0802.0762}{{\tt
  0802.0762}}].

\bibitem{Depta:2020wmr}
P.~F. Depta, M.~Hufnagel, and K.~Schmidt-Hoberg, {\it {Robust cosmological
  constraints on axion-like particles}},  {\em JCAP} {\bf 05} (2020) 009,
  [\href{http://www.arxiv.org/abs/2002.08370}{{\tt 2002.08370}}].

\bibitem{Dev:2019hho}
P.~S.~B. Dev, R.~N. Mohapatra, and Y.~Zhang, {\it {Constraints on long-lived
  light scalars with flavor-changing couplings and the KOTO anomaly}},  {\em
  Phys. Rev. D} {\bf 101} (2020), no.~7 075014,
  [\href{http://www.arxiv.org/abs/1911.12334}{{\tt 1911.12334}}].

\bibitem{Foroughi-Abari:2020gju}
S.~Foroughi-Abari and A.~Ritz, {\it {LSND Constraints on the Higgs Portal}},
  {\em Phys. Rev.} {\bf D102} (2020), no.~3 035015,
  [\href{http://www.arxiv.org/abs/2004.14515}{{\tt 2004.14515}}].

\bibitem{Boiarska:2019jym}
I.~Boiarska, K.~Bondarenko, A.~Boyarsky, V.~Gorkavenko, M.~Ovchynnikov, and
  A.~Sokolenko, {\it {Phenomenology of GeV-scale scalar portal}},  {\em JHEP}
  {\bf 11} (2019) 162, [\href{http://www.arxiv.org/abs/1904.10447}{{\tt
  1904.10447}}].

\bibitem{Winkler:2018qyg}
M.~W. Winkler, {\it {Decay and detection of a light scalar boson mixing with
  the Higgs boson}},  {\em Phys. Rev. D} {\bf 99} (2019), no.~1 015018,
  [\href{http://www.arxiv.org/abs/1809.01876}{{\tt 1809.01876}}].

\bibitem{Ahn:2018mvc}
{\bf KOTO} {\bf Collaboration}, J.~K. Ahn {\em et~al.}, {\it {Search for the
  $K_L \!\to\! \pi^0 \nu \overline{\nu}$ and $K_L \!\to\! \pi^0 X^0$ decays at
  the J-PARC KOTO experiment}},  {\em Phys. Rev. Lett.} {\bf 122} (2019), no.~2
  021802, [\href{http://www.arxiv.org/abs/1810.09655}{{\tt 1810.09655}}].

\bibitem{Aaij:2015tna}
{\bf LHCb} {\bf Collaboration}, R.~Aaij {\em et~al.}, {\it {Search for
  hidden-sector bosons in $B^0 \!\to K^{*0}\mu^+\mu^-$ decays}},  {\em Phys.
  Rev. Lett.} {\bf 115} (2015), no.~16 161802,
  [\href{http://www.arxiv.org/abs/1508.04094}{{\tt 1508.04094}}].

\bibitem{Aaij:2016qsm}
{\bf LHCb} {\bf Collaboration}, R.~Aaij {\em et~al.}, {\it {Search for
  long-lived scalar particles in $B^+ \to K^+ \chi (\mu^+\mu^-)$ decays}},
  {\em Phys. Rev. D} {\bf 95} (2017), no.~7 071101,
  [\href{http://www.arxiv.org/abs/1612.07818}{{\tt 1612.07818}}].

\bibitem{Dev:2020eam}
P.~S.~B. Dev, R.~N. Mohapatra, and Y.~Zhang, {\it {Revisiting supernova
  constraints on a light CP-even scalar}},  {\em JCAP} {\bf 2008} (2020) 003,
  [\href{http://www.arxiv.org/abs/2005.00490}{{\tt 2005.00490}}].

\bibitem{merger}
P. S. B. Dev, J.-F. Fortin, S. P. Harris, K. Sinha, Y Zhang, in preparation.

\bibitem{Dev:2017ftk}
P.~S.~B. Dev, R.~N. Mohapatra, and Y.~Zhang, {\it {Lepton Flavor Violation
  Induced by a Neutral Scalar at Future Lepton Colliders}},  {\em Phys. Rev.
  Lett.} {\bf 120} (2018), no.~22 221804,
  [\href{http://www.arxiv.org/abs/1711.08430}{{\tt 1711.08430}}].

\bibitem{Cherchiglia:2017uwv}
A.~Cherchiglia, D.~St\"ockinger, and H.~St\"ockinger-Kim, {\it {Muon g-2 in the
  2HDM: maximum results and detailed phenomenology}},  {\em Phys. Rev. D} {\bf
  98} (2018) 035001, [\href{http://www.arxiv.org/abs/1711.11567}{{\tt
  1711.11567}}].

\bibitem{Dev:2018upe}
P.~S.~B. Dev, R.~N. Mohapatra, and Y.~Zhang, {\it {Probing TeV scale origin of
  neutrino mass at future lepton colliders via neutral and doubly-charged
  scalars}},  {\em Phys. Rev. D} {\bf 98} (2018), no.~7 075028,
  [\href{http://www.arxiv.org/abs/1803.11167}{{\tt 1803.11167}}].

\bibitem{Chun:2019oix}
E.~J. Chun, J.~Kim, and T.~Mondal, {\it {Electron EDM and Muon anomalous
  magnetic moment in Two-Higgs-Doublet Models}},  {\em JHEP} {\bf 12} (2019)
  068, [\href{http://www.arxiv.org/abs/1906.00612}{{\tt 1906.00612}}].

\bibitem{Wang:2021fkn}
H.-X. Wang, L.~Wang, and Y.~Zhang, {\it {muon $g-2$ anomaly and
  $\mu$-$\tau$-philic Higgs doublet with a light CP-even component}},
  \href{http://www.arxiv.org/abs/2104.03242}{{\tt 2104.03242}}.

\bibitem{Kainulainen:2015sva}
K.~Kainulainen, K.~Tuominen, and V.~Vaskonen, {\it {Self-interacting dark
  matter and cosmology of a light scalar mediator}},  {\em Phys. Rev. D} {\bf
  93} (2016), no.~1 015016, [\href{http://www.arxiv.org/abs/1507.04931}{{\tt
  1507.04931}}]. [Erratum: Phys.Rev.D 95, 079901 (2017)].

\bibitem{Fradette:2017sdd}
A.~Fradette and M.~Pospelov, {\it {BBN for the LHC: constraints on lifetimes of
  the Higgs portal scalars}},  {\em Phys. Rev. D} {\bf 96} (2017), no.~7
  075033, [\href{http://www.arxiv.org/abs/1706.01920}{{\tt 1706.01920}}].

\bibitem{Dev:2017dui}
P.~S.~B. Dev, R.~N. Mohapatra, and Y.~Zhang, {\it {Long Lived Light Scalars as
  Probe of Low Scale Seesaw Models}},  {\em Nucl. Phys. B} {\bf 923} (2017)
  179--221, [\href{http://www.arxiv.org/abs/1703.02471}{{\tt 1703.02471}}].

\bibitem{Ishizuka:1989ts}
N.~Ishizuka and M.~Yoshimura, {\it {Axion and Dilaton Emissivity From Nascent
  Neutron Stars}},  {\em Prog. Theor. Phys.} {\bf 84} (1990) 233--250.

\bibitem{Hanhart:2000er}
C.~Hanhart, D.~R. Phillips, S.~Reddy, and M.~J. Savage, {\it {Extra dimensions,
  SN1987a, and nucleon-nucleon scattering data}},  {\em Nucl. Phys. B} {\bf
  595} (2001) 335--359, [\href{http://www.arxiv.org/abs/nucl-th/0007016}{{\tt
  nucl-th/0007016}}].

\bibitem{Arndt:2002yg}
D.~Arndt and P.~J. Fox, {\it {Saxion emission from SN1987A}},  {\em JHEP} {\bf
  02} (2003) 036, [\href{http://www.arxiv.org/abs/hep-ph/0207098}{{\tt
  hep-ph/0207098}}].

\bibitem{Diener:2013xpa}
R.~Diener and C.~P. Burgess, {\it {Bulk Stabilization, the Extra-Dimensional
  Higgs Portal and Missing Energy in Higgs Events}},  {\em JHEP} {\bf 05}
  (2013) 078, [\href{http://www.arxiv.org/abs/1302.6486}{{\tt 1302.6486}}].

\bibitem{Krnjaic:2015mbs}
G.~Krnjaic, {\it {Probing Light Thermal Dark-Matter With a Higgs Portal
  Mediator}},  {\em Phys. Rev.} {\bf D94} (2016), no.~7 073009,
  [\href{http://www.arxiv.org/abs/1512.04119}{{\tt 1512.04119}}].

\bibitem{Lee:2018lcj}
J.~S. Lee, {\it {Revisiting Supernova 1987A Limits on Axion-Like-Particles}},
  \href{http://www.arxiv.org/abs/1808.10136}{{\tt 1808.10136}}.

\bibitem{Araki:2020wkq}
T.~Araki, K.~Asai, H.~Otono, T.~Shimomura, and Y.~Takubo, {\it {Dark Photon
  from Light Scalar Boson Decays at FASER}},  {\em JHEP} {\bf 03} (2021) 072,
  [\href{http://www.arxiv.org/abs/2008.12765}{{\tt 2008.12765}}].

\bibitem{Zhang:2014wra}
Y.~Zhang, {\it {Supernova Cooling in a Dark Matter Smog}},  {\em JCAP} {\bf
  1411} (2014) 042, [\href{http://www.arxiv.org/abs/1404.7172}{{\tt
  1404.7172}}].

\bibitem{Donoghue:1990xh}
J.~F. Donoghue, J.~Gasser, and H.~Leutwyler, {\it {The Decay of a Light Higgs
  Boson}},  {\em Nucl. Phys.} {\bf B343} (1990) 341--368.

\bibitem{Djouadi:2005gi}
A.~Djouadi, {\it {The Anatomy of electro-weak symmetry breaking. I: The Higgs
  boson in the standard model}},  {\em Phys. Rept.} {\bf 457} (2008) 1--216,
  [\href{http://www.arxiv.org/abs/hep-ph/0503172}{{\tt hep-ph/0503172}}].

\bibitem{Zyla:2020zbs}
{\bf Particle Data Group} {\bf Collaboration}, P.~A. Zyla {\em et~al.}, {\it
  {Review of Particle Physics}},  {\em PTEP} {\bf 2020} (2020), no.~8 083C01.

\bibitem{Krnjaic:2019rsv}
G.~Krnjaic, G.~Marques-Tavares, D.~Redigolo, and K.~Tobioka, {\it {Probing
  Muonphilic Force Carriers and Dark Matter at Kaon Factories}},  {\em Phys.
  Rev. Lett.} {\bf 124} (2020), no.~4 041802,
  [\href{http://www.arxiv.org/abs/1902.07715}{{\tt 1902.07715}}].

\bibitem{Ecker:1987qi}
G.~Ecker, A.~Pich, and E.~de~Rafael, {\it {$K \to \pi l^+ l^-$ Decays in the
  Effective Chiral Lagrangian of the Standard Model}},  {\em Nucl. Phys.} {\bf
  B291} (1987) 692--719.

\bibitem{DAmbrosio:1998gur}
G.~D'Ambrosio, G.~Ecker, G.~Isidori, and J.~Portoles, {\it {The Decays $K \to
  \pi l^+ l^-$ beyond leading order in the chiral expansion}},  {\em JHEP} {\bf
  08} (1998) 004, [\href{http://www.arxiv.org/abs/hep-ph/9808289}{{\tt
  hep-ph/9808289}}].

\bibitem{Pospelov:2008zw}
M.~Pospelov, {\it {Secluded U(1) below the weak scale}},  {\em Phys. Rev.} {\bf
  D80} (2009) 095002, [\href{http://www.arxiv.org/abs/0811.1030}{{\tt
  0811.1030}}].

\bibitem{Davoudiasl:2014kua}
H.~Davoudiasl, H.-S. Lee, and W.~J. Marciano, {\it {Muon $g-2$, rare kaon
  decays, and parity violation from dark bosons}},  {\em Phys. Rev.} {\bf D89}
  (2014), no.~9 095006, [\href{http://www.arxiv.org/abs/1402.3620}{{\tt
  1402.3620}}].

\bibitem{Adams:2013qkq}
{\bf LBNE} {\bf Collaboration}, C.~Adams {\em et~al.}, {\it {The Long-Baseline
  Neutrino Experiment: Exploring Fundamental Symmetries of the Universe}},  in
  {\em {Snowmass 2013}: {Workshop on Energy Frontier}}, 7, 2013.
\newblock \href{http://www.arxiv.org/abs/1307.7335}{{\tt 1307.7335}}.

\bibitem{Ahn:2020opg}
{\bf KOTO} {\bf Collaboration}, J.~K. Ahn {\em et~al.}, {\it {Study of the $K_L
  \!\to\! \pi^0 \nu \overline{\nu}$ Decay at the J-PARC KOTO Experiment}},
  {\em Phys. Rev. Lett.} {\bf 126} (2021), no.~12 121801,
  [\href{http://www.arxiv.org/abs/2012.07571}{{\tt 2012.07571}}].

\bibitem{Ohl:1990qw}
K.~Ohl {\em et~al.}, {\it {A Measurement of the Branching Ratio and Form-factor
  for $K_L \to e^+ e^- \gamma$}},  {\em Phys. Rev. Lett.} {\bf 65} (1990)
  1407--1410.

\bibitem{Barr:1990zh}
{\bf NA31} {\bf Collaboration}, G.~Barr {\em et~al.}, {\it {Measurement of the
  Rate of the Decay $K_L \to e^+ e^- \gamma$ and Observation of a Form-factor
  in This Decay}},  {\em Phys. Lett. B} {\bf 240} (1990) 283--288.

\bibitem{Fanti:1999rz}
{\bf NA48} {\bf Collaboration}, V.~Fanti {\em et~al.}, {\it {Measurement of the
  decay rate and form-factor parameter $\alpha(K^*)$ in the decay $K_L \to e^+
  e^- \gamma$}},  {\em Phys. Lett. B} {\bf 458} (1999) 553--563.

\bibitem{AlaviHarati:2001wd}
{\bf KTeV} {\bf Collaboration}, A.~Alavi-Harati {\em et~al.}, {\it {Measurement
  of the branching ratio and form-factor of $K_L \to \mu^+ \mu^- \gamma$}},
  {\em Phys. Rev. Lett.} {\bf 87} (2001) 071801.

\end{thebibliography}\endgroup

\end{document}